\documentclass[article]{elsarticle}
\journal{Physical Review  D}
\newcommand{\dint}{{\rm d}}
  \usepackage{amsmath}
\usepackage{cleveref}
\usepackage{amsfonts}
\usepackage{graphicx}% Include figure files
\usepackage{dcolumn}% Align table columns on decimal point
\usepackage{bm}% bold math
\usepackage{caption}
\usepackage{subcaption}
\usepackage{amssymb}

\begin{document}
\begin{frontmatter}

\title{Investigating saturation effects and the virtual pion in leading neutron events at HERA with the dipole model}

\author[myaddress]{Arjun Kumar \corref{correspondingauthor}}
\address[myaddress]{Department of Physics, Indian Institute of Technology Delhi, India}
\cortext[correspondingauthor]{Corresponding author}
\ead{arjun.kumar@physics.iitd.ac.in}
\author[myaddress]{Tobias Toll}
\begin{abstract}
We investigate events with very forward neutrons in $ep$ collisions at HERA using impact parameter dependent colour dipole models with and without saturation. This is the first study of the leading neutron process deploying these models. The model predictions are compared with the available HERA measurements for $6<Q^2<100$~GeV$^2$ ,  $70<W<245$~GeV. Our analysis shows that the models exhibit Feynman scaling, independent of $Q^2$. Our results demonstrate that the $W$ and $Q^2$  dependence of the cross section is independent of the presence of a forward neutron as predicted by limiting fragmentation hypothesis, which is a consequence of Feynman scaling itself. We infer that the HERA leading neutron production inclusive data is insensitive to saturation physics and these cross sections may not be able to distinguish gluon saturation effects in future $ep$ colliders. We provide a good description of the leading neutron structure function $F_2^{LN}$ at small $x$ using an assumption that the small-$x$ structure of protons and pions is universal up to a normalisation. We also show that the observables in the exclusive diffractive measurements with a vector meson in the final state are more sensitive to saturation physics at small $x$ than inclusive measurements. At last we provide a prediction for the $\hat t$ spectrum in exclusive vector meson production in the dipole model using Yukawa theory to model the virtual pion's spatial wave function.
\end{abstract}

\begin{keyword}
Feynman scaling\sep Leading neutrons \sep Saturation
\end{keyword}

\end{frontmatter}

%\linenumbers

\section{Introduction}
 The colour dipole model provides an unified framework to study the inclusive, exclusive, and inclusive diffractive data \cite{GolecBiernat:1998js,  GolecBiernat:1999qd,  Kowalski:2003hm,  Kowalski:2006hc,  Marquet:2007nf,  Kowalski:2008sa,  Rezaeian:2012ji,  Rezaeian:2013tka,  Mantysaari:2018nng,  Sambasivam:2019gdd} gathered in \emph{ep} collisions and the inclusive  particle production data in \emph{pp}, \emph{p}A, and AA collisions \cite{Levin:2011hr,Tribedy:2010ab,Tribedy:2011aa} at small \emph{x}. Recently this framework has been extended to study the production of neutrons in very forward direction which carry a large fraction of longitudinal momentum ($x_{L} > 0.3$) of the protons in \emph{ep} collisions \cite{Goncalves:2015mbf,Carvalho:2015eia, Amaral:2018xsm,Carvalho:2020myz}. These are usually known as the leading neutrons. In the dipole picture, formulated in the target's rest frame, the virtual photon emitted from the incoming electron splits into quark-antiquark pair forming a colour dipole which subsequently interacts with the target. In the case of leading neutrons, the dipole probes the pion cloud of the proton, and the forward neutron comes from the proton as it splits into a neutron and a positive pion. 
%{\comment: put later?}

Leading neutron production has been extensively studied at the HERA \emph{ep} collider experiments H1 and ZEUS  \cite{H1:2010hym,ZEUS:2002gig}. Recently this data has been used to constrain the gluon density function of pions at small-$x$ in a global QCD analysis \cite{Barry:2018ort,Barry:2021osv}. The H1 collaboration performed the measurements of the Feynman-\emph{x} spectrum of the inclusive leading neutrons for photon virtualities $6<Q^2<100~$GeV$^2$ , and photon-proton centre of mass energies $70<W<245~$GeV \cite{H1:2014psx} and found the data to be in agreement with the Feynman scaling \cite{PhysRevLett.23.1415} and the limiting fragmentation hypothesis \cite{PhysRev.188.2159}, which for this case means that the $x_L$ spectrum of the interaction is independent of $Q^2$ and $W$. For small $Q^2$, the dipole size is large and it can re-scatter by interacting with the final sate neutron giving rise to absorptive corrections. The H1 measurements show that these absorptive corrections are sizeable in the small $Q^2$ region. These corrections have been recently calculated in \cite{Carvalho:2020myz} where the authors demonstrated that the absorptive effects are not strongly energy dependent and can be modelled by a $Q^2$ dependent multiplicative factor. 

Earlier attempts by Carvalho, Gon\c{c}alves, Spiering, and Navarra (CGSN) \cite{Carvalho:2015eia} to explain the Feynman scaling observed at HERA in the leading neutron spectrum showed that this scaling is associated with gluon saturation and only exists for small $Q^2$ values near the saturation scale $Q_S^2 \sim 1-2~$ GeV$^2$. CGSN argue, by using the so called bCGC dipole model \cite{Kowalski:2006hc, Rezaeian:2013tka}, that this scaling is due to the saturation of the dipole cross section at higher energies. This is surprising, as saturation is expected to only become prominent at small $x$, and the $x$ values probed in semi-inclusive measurements of leading neutrons is considerably larger than what has been probed in inclusive DIS, where the latter has exhibited no clear signal for saturation. 
% at HERA kinematics in any of the dipole models, also both the models with and without saturation explains the HERA $F_2$ data very well and there is no distinct signature of saturation in the data. 
Further in \cite{Mantysaari:2018nng} the authors showed that the structure function $F_2$ will be insensitive to saturation effects even in the kinematic region of future \emph{ep} colliders such as the FCC, or the LHeC \cite{Agostini:2020fmq}.  Moreover, there is also a scaling with respect to $Q^2$ in the leading neutron cross section observed in the HERA measurements for which the scaling with respect to $W$ in the Feynman-$x$ spectrum should be present for all $Q^2$ values. This raises more concerns on whether or not saturation effects lead to Feynman scaling. 

This paper aims at investigating these intriguing questions. We use two versions of the impact-parameter dependent dipole model, one which saturates at large dipoles and small $x$, named bSat (or IP-Sat), and a linearised version without saturation named bNonSat \cite{Kowalski:2006hc, Rezaeian:2012ji, Mantysaari:2018nng,  Sambasivam:2019gdd}. This is the first time these models are used to describe leading neutron data. One benefit of the bSat model compared to the bCGC model is that the former contains an explicit transverse profile of the target which we will utilise in this paper. We will demonstrate that the Feynman scaling is not associated with saturation in the kinematic regime accessible at HERA or a future Electron-Ion collider \cite{Accardi:2012qut,AbdulKhalek:2021gbh}, FCC or LHeC. Rather, this is a consequence of the identical asymptotic behaviour of the pion structure function, $F_2^{\pi}$, and the proton structure function $F_2$ at small \emph{x}. We observe that the leading neutron cross section has a scaling with respect to both $Q^2$ and $W$ in our models. 

We further investigate whether saturation effects can be seen in leading neutron data with exclusively produced vector mesons. %The models based on the dipole picture are parameterisations which are fitted to inclusive HERA data. 
We propose a novel way to calculate the $\hat t$ spectrum of exclusive vector meson production with a leading neutron, using Yukawa theory, and present our predictions. For this observable the universality in the gluon structure between protons and pions is expected to break down, as the shape of the $\hat t$ spectrum in $\gamma^*\pi^*$ collisions will be distinctly different from the $t$ spectrum in $\gamma^*p$ collisions.

The paper is organised as follows. In the next section we give a brief outline of the leading neutron production in the dipole picture and discuss the necessary ingredients to calculate the differential cross sections of leading neutrons. In section \ref{results}, we present our results on the scaling of the cross sections with respect to $W$ and $Q^2$  in the presence of a forward neutron and compare our predictions with available HERA data. We also provide an estimate of the saturation effects and the $t$-dependence in an exclusive measurement of leading neutrons. In the end, we summarise and discuss the main conclusions of our study.

%%%%%%%%%%%%%%%%%%%%%%%%%%%%%%%%%%%%%%%%%%%%%%%%%%%
\section{Leading neutrons in the dipole model}
\begin{figure}
	\centering
	\begin{subfigure}[b]{0.45\textwidth}
		\centering
		\includegraphics[width=0.85\linewidth]{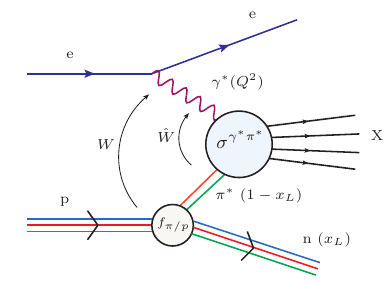}
			\caption{}
	\end{subfigure}
	\hfill
	\begin{subfigure}[b]{0.45\textwidth}
		\centering
 \includegraphics[width=0.9\linewidth]{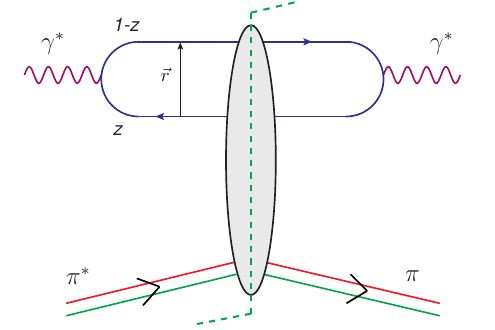}
 	\caption{}
	\end{subfigure}
	\caption{Leading neutron production in one-pion exchange approximation in \emph{ep} collisions (a) and $\gamma^*\pi^*  $  scattering cross section in dipole model (b)}
	\label{Lneutron_dipole}
\end{figure}

%%%%%%%%%%%%%%%%%%%%%%%%%%%%%%%%%%%%%%%%%%%%%%%%%%%
\subsection{The leading neutron structure function $F_2^{LN}$}
 The differential cross section for inclusive $ep\rightarrow e X$ scattering is related to the proton structure function $F_2(x,Q^2)$ as follows:
\begin{equation}
\frac{\dint^2\sigma^{ep\rightarrow e X}}{\dint x\dint Q^2} = \frac{4 \pi \alpha_{\rm EM}^2}{x Q^4} \biggl(1-y+\frac{y^2}{2}\biggr)F_2(x,Q^2)
\end{equation}
where $y$ is the virtual photon elasticity, $Q^2$ its virtuality, $x$ the momentum fraction of the proton's longitudinal momentum taken by the struck parton, and $\alpha_{\rm EM}$ is the electro-magnetic coupling. We may write the structure function in terms of the $\gamma^*p$ cross section as:
\begin{equation}
F_2 = \frac{ Q^2}{4 \pi^2 \alpha_{\rm EM}}(\sigma_L^{\gamma^*p}+\sigma_T^{\gamma^*p})
\end{equation}
 the total $\gamma^*p$ cross section in the dipole model is calculated using the optical theorem which states that the cross section is the imaginary part of the forward elastic $\gamma^*p \rightarrow \gamma^*p$ amplitude and is given by \cite{Kowalski:2006hc}:
 \begin{equation}
\sigma_{L,T}^{\gamma^*p}(x,Q^2) = \int \dint^2\textbf{b}~\dint^2\textbf{r}\int_{0}^{1} \frac{\dint z}{4 \pi} |\Psi^f_{L,T}(\textbf{r},z,Q^2)|^2 \frac{\dint\sigma^{(p)}_{q\bar{q}}}{\dint^2\textbf{b}}(\textbf{b},\textbf{r},x)
\end{equation}
with $z$ being the photon's momentum fraction taken by the quark, $\bf{r}$ the dipole's transverse size and direction, and ${\bf b}$ the impact parameter. Analogous to this, for leading neutrons we can write the differential cross section for $ep\rightarrow e X n$, depicted in Fig.~\ref{Lneutron_dipole}(a), as \cite{H1:2010hym,ZEUS:2002gig}: 
\begin{equation}
\frac{\dint^4\sigma^{ep\rightarrow e Xn}}{\dint x\dint Q^2\dint x_L\dint t} = \frac{4 \pi \alpha_{\rm EM}^2}{x Q^4} \biggl(1-y+\frac{y^2}{2}\biggr)F_2^{LN(4)}(x,Q^2,x_L,t).
\end{equation}
where $t$ is the four-momentum transfer squared at the proton vertex, $x_L$ is the proton's longitudinal momentum fraction taken by the neutron, while the pion takes $1-x_L$, as illustrated in Fig.~\ref{Lneutron_dipole}(b). For leading neutrons we get the following relation for $F_2^{LN}$ in terms of $\gamma^*p$ cross section as:
\begin{equation}
F_2^{LN}(x,Q^2,x_L)= \frac{ Q^2}{4 \pi^2 \alpha_{\rm EM}}\frac{\dint \sigma^{\gamma^*p\rightarrow Xn}}{\dint x_L}
\end{equation}
where,
\begin{equation}
\frac{\dint \sigma^{\gamma^*p\rightarrow Xn}}{\dint x_L} =  \int_{t_{min}}^{t_{max}}\frac{\dint^2 \sigma^{\gamma^*p\rightarrow Xn}}{\dint x_L\dint t } \dint t
\end{equation}
In the one-pion exchange approximation \cite{BISHARI1972510}, at high energies, the differential cross section for $\gamma^*p\rightarrow Xn$ can be written as:
\begin{equation}
\frac{\dint^2 \sigma (W,Q^2,x_L,t)}{\dint x_L \dint t } = f_{\pi/p}(x_L,t) ~\sigma^{\gamma^* \pi^*}(\hat{W^2},Q^2)
\end{equation}
where $ f_{\pi/p}(x_L,t)$ is the flux of pions emitted by the proton and $\sigma^{\gamma^* \pi^*}$ is the cross section of $\gamma^*\pi^*$ interactions. The leading neutron structure function becomes \cite{H1:2010hym,ZEUS:2002gig}:
\begin{eqnarray}
F_2^{LN}(W,Q^2,x_L)&= \frac{ Q^2}{4 \pi^2 \alpha_{\rm EM}}  \int_{t_{min}}^{t_{max}} f_{\pi/p}(x_L,t) ~ \sigma^{\gamma^* \pi^*}(\hat{W^2},Q^2) ~ \dint t\\
&=    \int_{t_{min}}^{t_{max}} f_{\pi/p}(x_L,t) ~\mathcal{K}(Q^2)~ F_2^{\pi}(W,Q^2,x_L)~ \dint t\\
&= \Gamma(x_L,Q^2) F_2^{\pi}(W,Q^2,x_L) 
\end{eqnarray}
where $\mathcal{K}(Q^2)$ is a multiplicative factor which includes the effect of absorptive corrections which in general modifies the flux. Here, $ F_2^{\pi}(W,Q^2,x_L) =  \frac{ Q^2}{4 \pi^2 \alpha_{\rm EM}}\sigma^{\gamma^*\pi^*}(\hat{W^2},Q^2) $ is the pion structure function and $\Gamma(x_L,Q^2) = \mathcal{K}(Q^2)\int_{t_{min}}^{t_{max}} $\\$f_{\pi/p}(x_L,t)~\dint t$ is the pion flux factor integrated over the \emph{t}-region of the measurement and corrected for the absorptive effects. Here, \emph{W} is the centre-of-mass energy for the photon-proton system, $\hat{W}$ is the centre-of-mass energy for the photon-pion system with $\hat{W}^2=(1-x_L)W^2$. The $t$ variable is related to $p_T$, the transverse momentum of the neutron, and $x_L$ as:
\begin{equation}
t \simeq -\frac{p_T^2}{x_L}-(1-x_L)\left(\frac{m_n^2}{x_L}-m_p^2\right)
\end{equation}
where $m_n$ and $m_p$ are the masses of neutron and proton respectively.
%%%%%%%%%%%%%%%%%%%%%%%%%%%%%%%%%%%%%%%%%%%%%%%%%%%%%%%
\subsection{The pion flux}
The flux factor $ f_{\pi/p}(x_L,t)$  describes the splitting of a proton into a $\pi n$ system. 
%In general we could use the flux from the light cone Fock state $\pi n$ of a proton from the meson-cloud models \cite{DAlesio:1998uav} or recall the expression from triple-Regge diagrams which describe inclusive processes as $ap \rightarrow Xn$ with $a = \gamma$ or $p$ \cite{Pumplin:1973ref,Kaidalov:2006cw,Nikolaev:1997cn}. 
This flux  parametrisation with dominant-pion exchange contribution has been used to explain hadron-hadron interactions and data from the H1 and ZEUS experiments \cite{ZEUS:2002gig,H1:2010hym,H1:2014psx} and in the earlier analysis in \cite{Goncalves:2015mbf,Carvalho:2015eia, Amaral:2018xsm,Carvalho:2020myz}. The flux factor is given by:
\begin{equation}
f_{\pi/p}(x_L,t)=\frac{1}{4 \pi}\frac{2 g^2_{p\pi p}}{4 \pi} \frac{|t|}{(m_\pi^2+|t|)^2}(1-x_L)^{1-2\alpha(t)}[F(x_L,t)]^2
\label{eq:flux}
\end{equation} 
where $m_\pi$ is the pion masss, $g^2_{p\pi p}/(4\pi)=14.4$ is the $\pi^0pp$ coupling. $F(x_L,t)$ is the form factor which accounts for the finite size of the vertex. This kind of splitting function $f_{\pi/p}$ can be also evaluated using chiral effective theory while the form factor introduces some model dependence. We consider the covariant form factor, corrected by a Regge factor for our analysis:
\begin{equation}
F(x_L,t) = \exp\bigg[-R^2\frac{|t|+m_\pi^2}{(1-x_L)}\bigg], \alpha(t) = 0
\end{equation}
where $R=0.6~$GeV$^{-1}$ has been determined from HERA data \cite{HOLTMANN1994363}.

%%%%%%%%%%%%%%%%%%%%%%%%%%%%%%%%%%%%%%%%%%%%%%%%%%%%%%%
\subsection{The total photon-pion cross section $\sigma^{\gamma^*\pi^*}$ in the dipole model}
Using the optical theorem, the total $\gamma^*\pi^*$ cross section is given by the imaginary part of the forward elastic $\gamma^*\pi^* \rightarrow \gamma^*\pi^*$ amplitude. We use the dipole model to calculate the $\gamma^* \pi^*$ cross section. In the dipole picture, at high energies, this  scattering amplitude factorises and is given by convolution of three subprocess, as depicted in Fig. 1. First, the virtual photon splits into a quark anti-quark dipole, then the dipole interacts with the pion via one or many gluon exchanges and then forms the final state which is a virtual photon in this case. Thus the total $\gamma^*\pi^*$ cross section is given by:
 \begin{equation}
\sigma_{L,T}^{\gamma^*\pi^*}(\hat{x},Q^2) = \int \dint^2\textbf{b}~\dint^2\textbf{r}\int_{0}^{1} \frac{\dint z}{4 \pi} |\Psi^f_{L,T}(\textbf{r},z,Q^2)|^2 \frac{\dint\sigma^{(\pi)}_{q\bar{q}}}{\dint^2\textbf{b}}(\textbf{b},\textbf{r},\hat{x})
\end{equation}
where $\hat{x}$ is the scaled Bjorken variable for the photon-pion system and is given by:
\begin{equation}
\hat{x} = \frac{Q^2+m_f^2}{\hat{W^2}+Q^2}= \frac{Q^2+m_f^2}{(1-x_L)W^2+Q^2}
\label{eq:x}
\end{equation}
The photon wavefunctions are well known quantities calculated in \cite{Kowalski:2006hc}. What remains to calculate in the total $\gamma^*\pi^*$ cross section is the dipole-pion cross section $\frac{\dint\sigma^{(\pi)}_{q\bar{q}}}{\dint^2\textbf{b}}$. We assume that the dipole-pion cross section is related to the dipole-proton cross section \cite{Kopeliovich:2012fd, Kaidalov:2006cw}, which has been determined through fits to inclusive HERA data. This means:
\begin{equation} 
\frac{\dint\sigma^{(\pi)}_{q\bar{q} }}{\dint^2\textbf{b}}(\textbf{b},\textbf{r},\hat{x}) = R_q~ \frac{\dint\sigma^{(p)}_{q\bar{q} }}{\dint^2\textbf{b}}(\textbf{b},\textbf{r},\hat{x})
\label{eq:scaling}
\end{equation}

This kind of assumption is also supported by the ZEUS analysis of the leading neutron data \cite{ZEUS:2002gig} where they showed that the proton structure function $F_2$ and the pion structure function $F_2^\pi$ are related. In a constituent quark picture, if we assume the additive quark model, then $R_q $ is the ratio of valence quarks in the pion and proton i.e. $R_q = \frac{2}{3}$. The same value of $R_q $ was obtained in a previous analysis of pion structure function at small \emph{x} based on the color dipole BFKL-Regge expansion in \cite{NIKOLAEV2000157} while the studies in \cite{PhysRevD.85.114025} concludes that this value could reach $R_q = 0.5$. We let $R_q$ vary in our study which provides an uncertainty band pertaining to different choices of $R_q$. It should be noted that according to eq.\eqref{eq:scaling}, apart from normalisation, the energy dependence of the pion structure function is identical to that of the proton at small \emph{x}. We will see below that this assumption is well justified. As a consequence of this, the pion structure function $F_2^\pi$ has the same asymptotic behaviour as the proton structure function $F_2$ in our models. We may also note that the pion is probed at larger $x$ than the proton, as $\hat x\geq x$ (with equality for  $x_L=0$).

We consider two versions of the dipole-proton cross section.
% we consider two versi of two versions of the dipole model: the saturated bSat model, which tames the growth the gluon density at small-$x$ and large $r$ by multiple two-gluon scatterings, and its linearised version the bNonSat model which describes a simple two gluon exchange.
The bSat model is given by:
\begin{eqnarray}
\frac{\dint\sigma^{(p)}_{q\bar{q}}}{\dint^2\textbf{b}}(\textbf{b},\textbf{r},x)=
2\big[1-\text{exp}\big(-F(x , r^2)T_p(\textbf{b})\big)\big]
\end{eqnarray}
with
\begin{eqnarray}
F(x , r^2)=\frac{\pi^2}{2N_C} r^2 \alpha_s(\mu^2) xg(x,\mu^2),
\end{eqnarray}

Due to the exponential functional form in this case, the dipole cross section saturates for large gluon density $xg(x, \mu^2)$  and for large dipole sizes $r$. 
The scale at which the strong coupling $\alpha_s$ and gluon density is evaluated at is $\mu^2 = \mu_0^2 +\frac{C}{r^2}$ and the gluon density at the initial scale $\mu_0$ is parametrised as:
\begin{eqnarray*}
	x g(x,\mu_0^2)= A_g x^{-\lambda_g}(1-x)^{6}
\end{eqnarray*}
The bNonSat model is a linearised version of bSat model where :
\begin{equation}
\frac{\dint\sigma^{(p)} _{q\bar{q}}}{\dint^2\textbf{b}}(\textbf{b},r,x)=\frac{\pi^2}{N_C}r^2\alpha_s(\mu^2) x g(x,\mu^2)  T_p(\textbf{b})
\end{equation}
which does not saturate for large gluon densities and large dipoles. The parameters $A_g, \lambda_g, C, m_f$ are determined through fits to the reduced cross section measured at HERA. We use the fit results from \cite{Sambasivam:2019gdd} where both models have been fitted independently. 

The transverse profile of the proton is assumed to be Gaussian:
\begin{eqnarray}
T_p(\textbf{b}) = \frac{1}{2 \pi B_p}\exp\bigg(-\frac{\textbf{b}^2}{2B_p}\bigg)
\label{eq:profile}
\end{eqnarray}
%\begin{figure}[h]
%	\centering
%	\includegraphics[width=0.65\linewidth]{Plots/F2_x_bSat_STU.pdf}
%	\caption{Proton structure function $F_2(x,Q^2)$ as function of \emph{x}, for different values of $Q^2$, in the bSat (solid line) and the bNonSat (dotted line) dipole models. The experimental data are from H1 and ZEUS collaborations \cite{H1:2009pze}.
%	}
%	\label{F2_dipolemodels}
%\end{figure}
The inclusive DIS cross sections are taken at $t=0$ and are only dependent on the profile function at non-leading twists. Therefore, the parameter $B_p$ is constrained through a fit to the $t$-dependence of the exclusive J/$\psi$ production at HERA \cite{Kowalski:2006hc, Rezaeian:2012ji}, and is found to be $B_p=4\pm 0.4~$GeV$^{-2}$. The profile function of the proton and pion will be discussed in detail below in the context of exclusive diffraction.

It should be noted that there is only one free parameter, $R_q$, in our study as the rest are fixed by inclusive DIS data, and the values of the multiplicative factor $\mathcal{K}$ in the flux corresponding to absorptive corrections has been taken from \cite{Carvalho:2020myz} where it has been explicitly calculated using high-energy Glauber approximation \cite{DAlesio:1998uav}. %Now we have all the ingredients to calculate the leading spectra and we show our results in next section. 
The experimental data is normalised with respect to the inclusive DIS cross section in form of $\frac{1}{\sigma_{\rm DIS}}\frac{\dint \sigma}{\dint x_L}$ . We could in principle calculate this cross section $\sigma_{\rm DIS}$ using the dipole model but we instead calculate it using the fitted parametrisation of $F_2$ data from \cite{ADLOFF2001183} to avoid any bias. It is:
\begin{equation}
\sigma_{\rm DIS} = \frac{4 \pi^2 \alpha} { Q^2} \frac{c}{x^\beta}
\end{equation}
where c = 0.18, $\beta=d\cdot\ln(Q^2/\Lambda_0^2)$ with $d= 0.0481$, and $\Lambda_0=0.292~$GeV. We have checked that using this parametrisation and the dipole model to calculate the inclusive cross section yield the same results.

Earlier studies such as \cite{Mantysaari:2018nng, Sambasivam:2019gdd} show that the $F_2$ structure function of the proton is insensitive to saturation effects. The authors conclude that even in the kinematic regime of future colliders such as the LHeC and the FCC, non-linear effects would be negligible for the proton structure function. For semi-inclusive measurements with leading neutrons, the probed $x$ value is higher as the available centre of mass energy for the virtual-pion photon system is given as $\hat{W^2}=(1-x_L)W^2$. Hence, it is expected that the leading neutron spectrum in $ep$ collisions will also be insensitive to non-linear effects. 
\begin{figure}
	\centering
	\begin{subfigure}[b]{0.45\textwidth}
		\centering
		\includegraphics[width=0.85\linewidth]{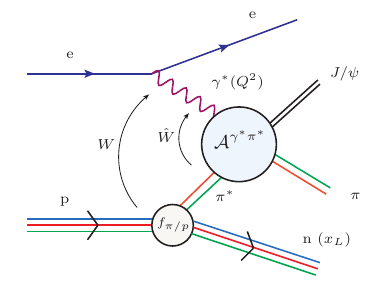}
		\caption{}
	\end{subfigure}
	\hfill
	\begin{subfigure}[b]{0.45\textwidth}
		\centering
		\includegraphics[width=0.9\linewidth]{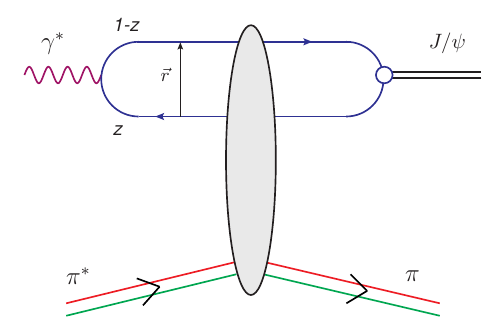}
		\caption{}
	\end{subfigure}
	
\caption{Exclusive J/$\psi$ production with leading neutron production in one-pion exchange approximation in \emph{ep} collisions (a) and $\gamma^*\pi ^* $  scattering amplitude in dipole model (b)}
	\label{Lneutron_dipole_Jpsi}
\end{figure}

%%%%%%%%%%%%%%%%%%%%%%%%%%%%%%%%%%%%%%%%%%%%%%%%%%%%%%
\subsection{Exclusive $J/\psi$ production  with leading neutron in the dipole model}
From investigations of $ep$ collisions at small $x$ at HERA, we know that for inclusive measurements the cross section is directly proportional to the gluon density while for diffractive measurements it is proportional to the gluon density squared, thus making the latter more sensitive to non-linear effects in the QCD evolution. Hence, the exclusive measurements of the Feynman-\emph{x} spectrum with leading neutrons and a vector meson in the final state has a greater potential for showing saturation effects.  For this, we study $e+p \rightarrow e'+J/\psi + \pi + n$ in \emph{ep} collisions where the vector meson is in the central detectors, while  the produced pion will disappear down the beam pipe, and the neutron can be measured by calorimeters in the very forward direction. We employ the one-pion exchange approximation to calculate the exclusive J/$\psi$ production with a leading neutron in the dipole framework as illustrated in Fig.~\ref{Lneutron_dipole_Jpsi}. The leading neutron cross section is in this case calculated as:
\begin{equation}
\frac{\dint^2 \sigma (W,Q^2,x_L,t)}{\dint x_L~\dint t } = f_{\pi/p}(x_L,t) ~\sigma^{\gamma^* \pi^*\rightarrow J/\psi~\pi}(\hat{W^2},Q^2)
\end{equation}
where $ f_{\pi/p}$ is the pion flux described in eq.\eqref{eq:flux}. 
The total $\gamma^*\pi^*$ cross section is given by \cite{Goncalves:2015mbf}:
\begin{equation}
\sigma^{\gamma^* \pi^*\rightarrow J/\psi~\pi} = \sum_{L,T}\int_{-\infty}^{0}\frac{\dint \sigma^{\gamma^* \pi^* \rightarrow J/\Psi~ \pi}}{\dint \hat{t}} \dint \hat{t} = \frac{1}{16 \pi}  \sum_{L,T}\int_{-\infty}^{0}\big| \mathcal{A}_{T,L}^{\gamma^* \pi^* \rightarrow J/\Psi ~\pi} \big|^2 \dint \hat{t}
\label{eq:exlusive}
\end{equation}
where the scattering amplitude is:
 \begin{eqnarray}
\mathcal{A}_{T,L}^{\gamma^* \pi^* \rightarrow J/\Psi ~\pi} (\hat{x},Q^2,\textbf{$\Delta$})=&i\int \dint^2 \textbf{r}\int \dint^2\textbf{b}\int \frac{\dint z}{4 \pi}   (\Psi^*\Psi_V)_{T,L}(Q^2,\textbf{r},z) \nonumber \\
&\times e^{-i[\textbf{b}-(1-z)\textbf{r}]\cdot\textbf{$\Delta$}} \frac{\dint\sigma^{(\pi)} _{q\bar{q}}}{\dint^2\textbf{b}}(\textbf{b},\textbf{r},\hat{x}).
\label{eqn:amplitude}
\end{eqnarray}
Here,  $(\Psi^*\Psi_V)$ is the wave-overlap of the photon and the vector-meson wave-functions. We use boosted the Gaussian wavefunction for $J/\psi$  with the parameter values from \cite{Mantysaari:2018nng}. The virtual pion dipole cross section $\dint\sigma^{(\pi)}_{q\bar{q}}/\dint^2\textbf{b}$ is given in eq.\eqref{eq:scaling}. This amplitude is a Fourier transform from coordinate space to momentum space, where $|\Delta|=\sqrt{-\hat t}$. The cross section is also corrected for skewedness and real correction (for details we refer to appendix of \cite{Mantysaari:2016jaz}). All the parameters that goes into our description of $J/\psi$ production with leading neutrons have thus been fixed by other processes. 
%We must emphasise that such a measurement would be direct evidence of the pion cloud in protons. 
One  of the greatest achievements of the pion-cloud models (see for e.g \cite{McKenney:2015xis}) is to explain the observed SU(2) flavour asymmetry ($\bar{d}-\bar{u}$ asymmetry, \cite{NuSea:2001idv}) at high-$x$ values. Hence this kind of exclusive measurement with a leading neutron would hint at the universality of pion flux from protons as well as  contributing to our understanding of the pion cloud in protons.

%%%%%%%%%%%%%%%%%%%%%%%%%%%%%%%%%%%%%%%%%%%%%%%%%%%%%%%
\subsection{Spatial distribution of gluons in exclusive vector meson production with leading neutrons} 
As discussed above, exclusive vector meson production with leading neutrons can be described in the one pion exchange approximation, where the dipole interacts with the virtual pion cloud of the proton. In such a measurement, the transverse momentum of the final state vector meson, $p_\perp$ can be measured in order to probe the spatial transverse gluon distribution in the virtual pion. We expect that the universality of the gluon distribution between protons and pions will break down in the $\hat t\approx p_\perp^2$ spectrum as they have different spatial profiles. This will not affect the inclusive observables discussed above, as they are calculated at $\hat t=0$.
%, neither will it affect the universality of cross sections in which the $\hat t$ distribution is integrated over. 

In order to calculate the differential cross section with respect to $\hat t$ we assume that the transverse profile of the virtual pion (the entire pion cloud) is given by a 2 dimensional Yukawa function: 
\begin{equation}
T_{\pi^*}(b)=\int_{-\infty}^\infty{\rm d}z\rho_{\pi^*}(b, z)
\label{eq:pionCloud}
\end{equation}
where the radial part of the virtual pion wave function is given by Yukawa theory:
\begin{eqnarray}
\rho_{\pi^*}(b, z)=\frac{m_\pi^2}{4\pi}\frac{e^{-m_\pi\sqrt{\textbf{b}^2+z^2}}}{\sqrt{\textbf{b}^2+z^2}}
\end{eqnarray}
We assume that the real pion, as for the proton, is described by a Gaussian profile:
\begin{eqnarray}
T_\pi(b)=\frac{1}{2\pi B_\pi}e^{-\frac{b^2}{2B_\pi}}
\end{eqnarray}
In the amplitude eq.~\eqref{eqn:amplitude} we see that $\Delta$ is the Fourier conjugate of $b$. We can interpret this as resolving the pion transverse wave-function at a spatial resolution $\delta b\sim 1/\Delta$. At small $|\hat t|$, the dipole interacts coherently with the entire virtual pion wave function, while at larger $|\hat t|$ the dipole will begin to resolve the pion $T_\pi$ inside the wave function $T_{\pi^*}$. This pion will have an event-by-event spatial distribution given by eq.~\eqref{eq:pionCloud}.
%We can thus model this in the Good-Walker framework \cite{Good:1960ba}, where the coherent cross section is given by $|\left<\mathcal{A}\right>|^2$, and the total cross section by $\left<|\mathcal{A}|^2\right>$, where the averaging $\big<\cdot\big>$ is over initial state positions. 
The total cross section in eq.~\eqref{eq:exlusive} is then given by:
\begin{eqnarray}
\left<|\mathcal{A}|^2\right>=|\left<\mathcal{A}\right>|^2+\left(\left<|\mathcal{A}|^2\right>-|\left<\mathcal{A}\right>|^2\right),
\end{eqnarray}
where the first term on the right hand side corresponds to the average position of the pion, and is given by $T_{\pi^*}$, while the second term is the event-by-event variation of the pion's position in the pion cloud and is achieved by sampling the positions of $T_\pi$ according to $T_{\pi^*}$. We thus calculate the differential cross section with respect to $\hat{t}$ by sampling the first and second moment of the real pion's position from the 2-dimensional Yukawa distribution.
%The variation in the cross section, given by $\left<|\mathcal{A}|^2\right>-|\left<\mathcal{A}\right>|^2$, comes from sampling real pions with a profile $T_\pi$ distributed as $T_{\pi^*}$.
%\footnote{In the Copenhagen interpretation, when the resolution is fine enough, the wave function $\rho_{\pi^*}$ collapses into the wave function of a real pion located at a specific impact parameter. For smaller $|\hat t|$, the real pion cannot be resolved.}
%hence the dipole probes the gluon distribution inside this region. Since the gluon distribution of the pion could be present anywhere in the region governed by eq. \eqref{pionCloud}, we have to sample the mean position of the gluons from eq.\eqref{pionCloud}. For simplicity we assume the profile to be a Gaussian distribution as follow: 
%where  $B_\pi$ is the width and $\textbf{b}_{\pi}$ is the mean position of then gluon distribution sampled from eq.\eqref{pionCloud}. In our formalism, we vary the mean positions event-by-event, hence this is equivalent to geometrical  fluctuations of the gluon distribution and we could employ Good-Walker framework to calculate the differential cross section as described in detail in \cite{Mantysaari:2016jaz,Mantysaari:2016ykx,Kumar:2022aly}.
%%%%%%%%%%%%%%%%%%%%%%%%%%%%%%%%%%%%%%%%%%%%%
\section{\label{results}Results}
\begin{figure*}[h]
	\centering
	\includegraphics[width=0.49\linewidth]{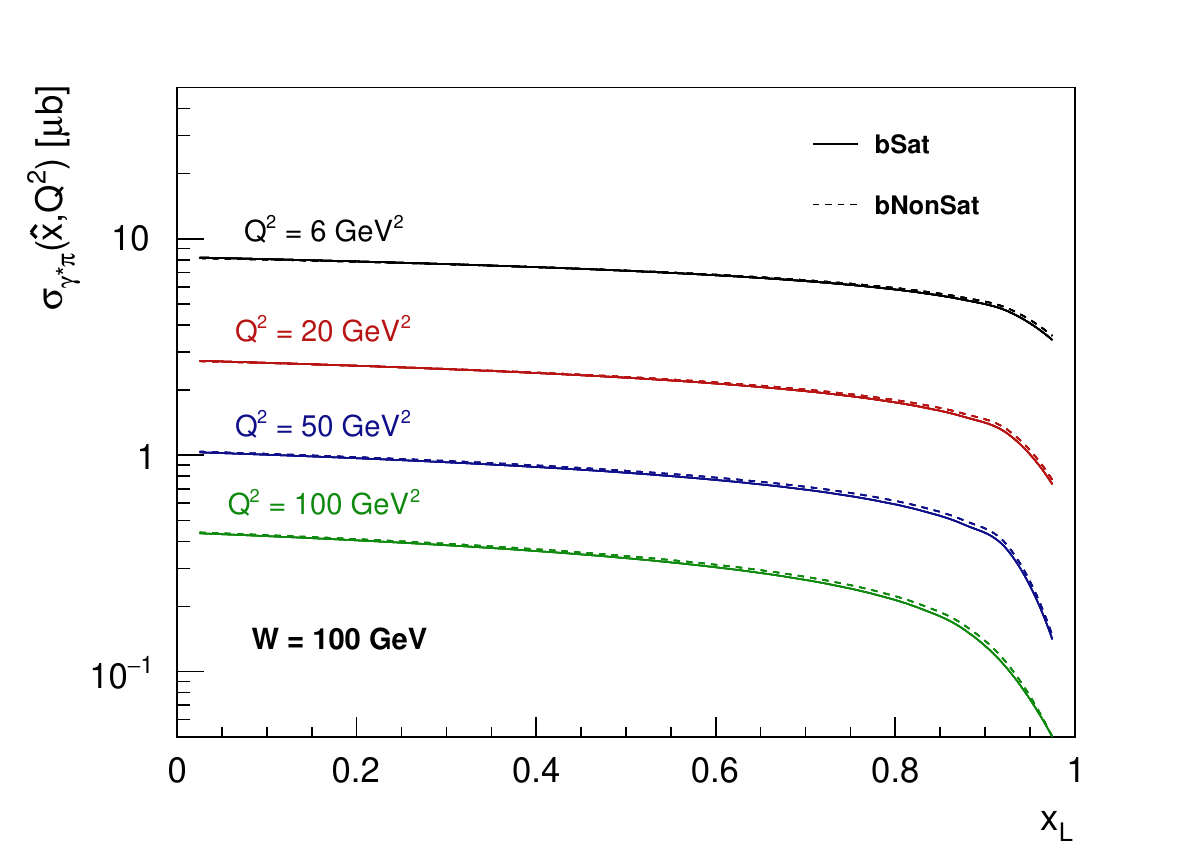}
	\includegraphics[width=0.49\linewidth]{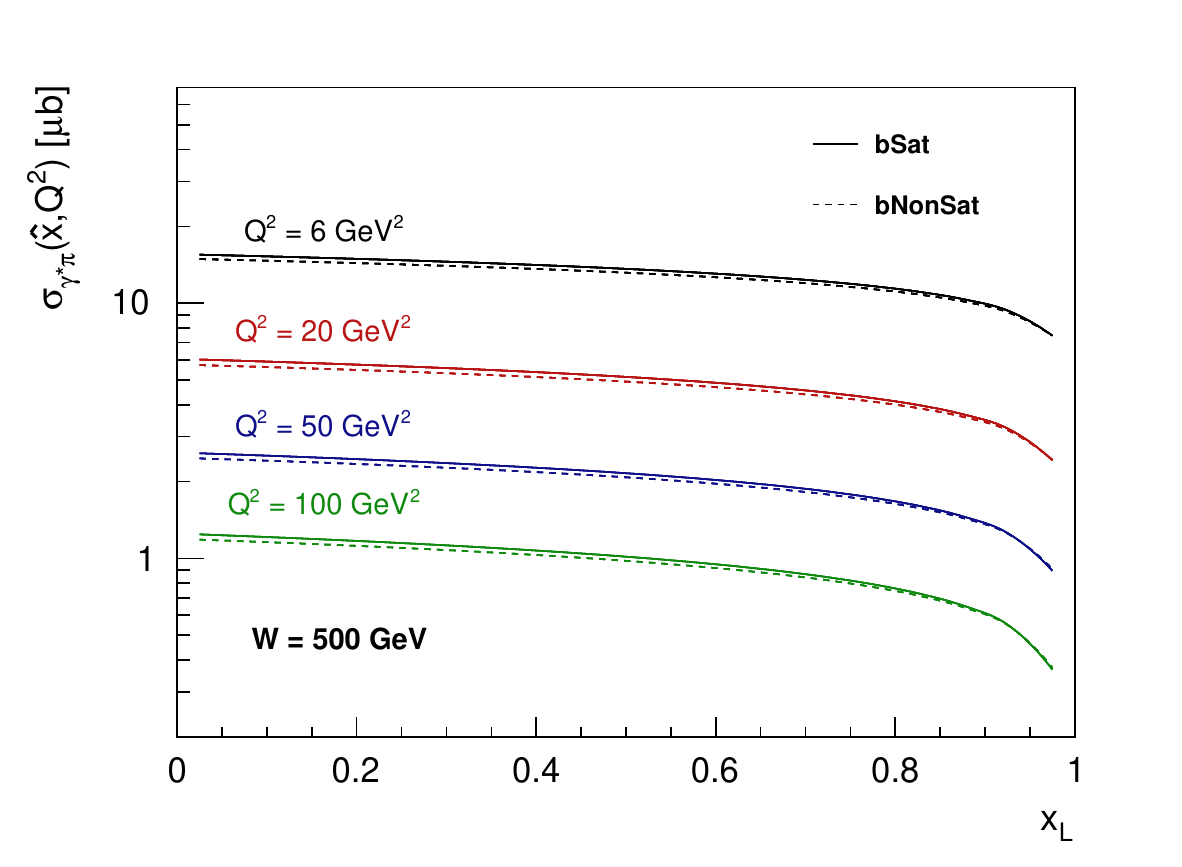}
	\includegraphics[width=0.49\linewidth]{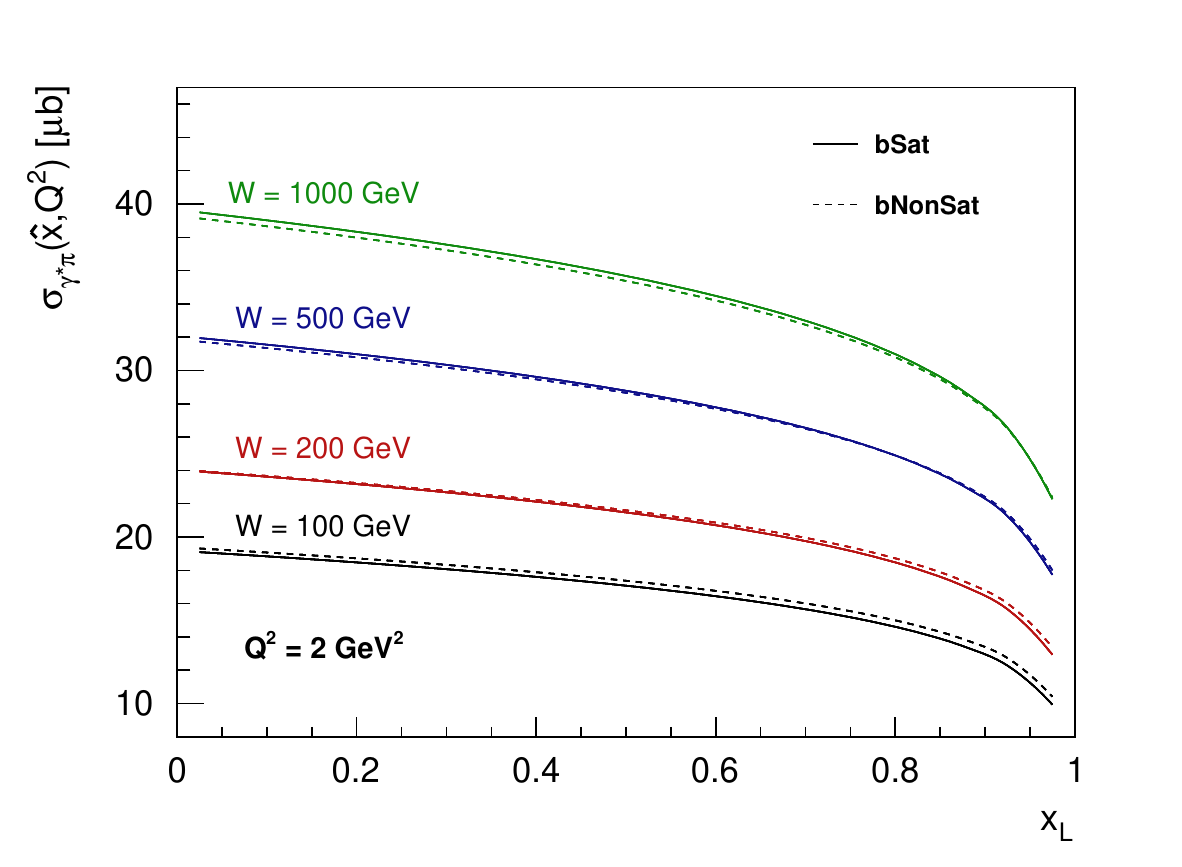}
	\includegraphics[width=0.49\linewidth]{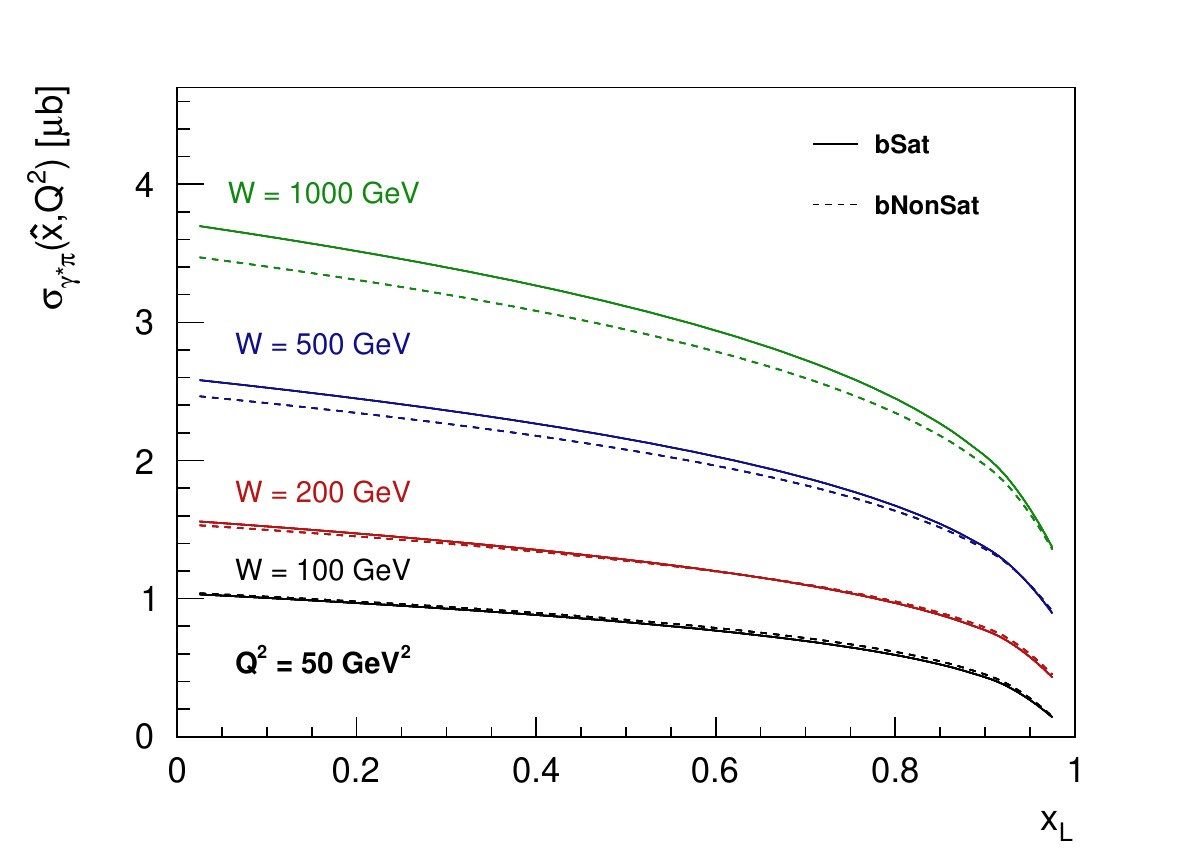}
	\caption{Dipole-pion cross section as a function of $x_L$ in bSat (solid line) and bNonSat(dashed line) dipole models with varying $Q^2$ in first row and with varying $W$ in second row }
	\label{dipole_pion}
\end{figure*}
%In Fig.~\ref{F2_dipolemodels}, we plot the proton structure function $F_2$ as a function of Bjorken-\emph{x} with varying $Q^2$ in bSat and the bNonSat dipole models. The curves for both the models are practically on top of each other in the whole kinematic region which show that no clear distinction can be made for the saturation effects to be present in HERA data. $F_2$ rises as we decrease \emph{x} in both the models which tells that the total $\gamma^*p$ cross section does not approaches a constant values for the studied \emph{x} regime even in the saturation model. We conclude from here that $F_2$ shows the same asymptotic behaviour in both with and without saturation model. This is similar to what was demonstrated in \cite{Mantysaari:2018nng} where the authors showed that  $F_2$ is be insensitive to non-linear physics and cannot distinguish saturation effects in the kinematic regime of present and future colliders like FCC or LHeC.   \\

In Fig.~\ref{dipole_pion}, we show the total photon-pion cross section as a function of longitudinal momentum fraction of the proton carried by the neutron, $x_L$, in the bSat and bNonSat dipole models. The cross section rises with $W$  for both small and large $Q^2$ in both models. As expected there is no saturation observed, since the  dipole-pion and dipole-proton cross sections are related and we have not seen any saturation effects in the latter. This is important as the photon-pion cross section is an integral part of the calculation of the semi-inclusive leading neutron spectrum which exhibits Feynman scaling, hence such a scaling, if it exists, should be found in both models. The dipole-pion cross section also decreases with the increasing values of $Q^2$ which is the expected high energy behaviour. It appears that for very large $W$ with $Q^2=50~$GeV$^2$ the bSat model is slighty above the bNonSat model. We have checked that this is an effect from extrapolating the model fits away from the kinematics of the inclusive measurements that are available to the fits, and is a result from differing parameters in the bNonSat and bSat models.

\begin{figure*}[h]
	\centering
	\includegraphics[width=0.5\linewidth]{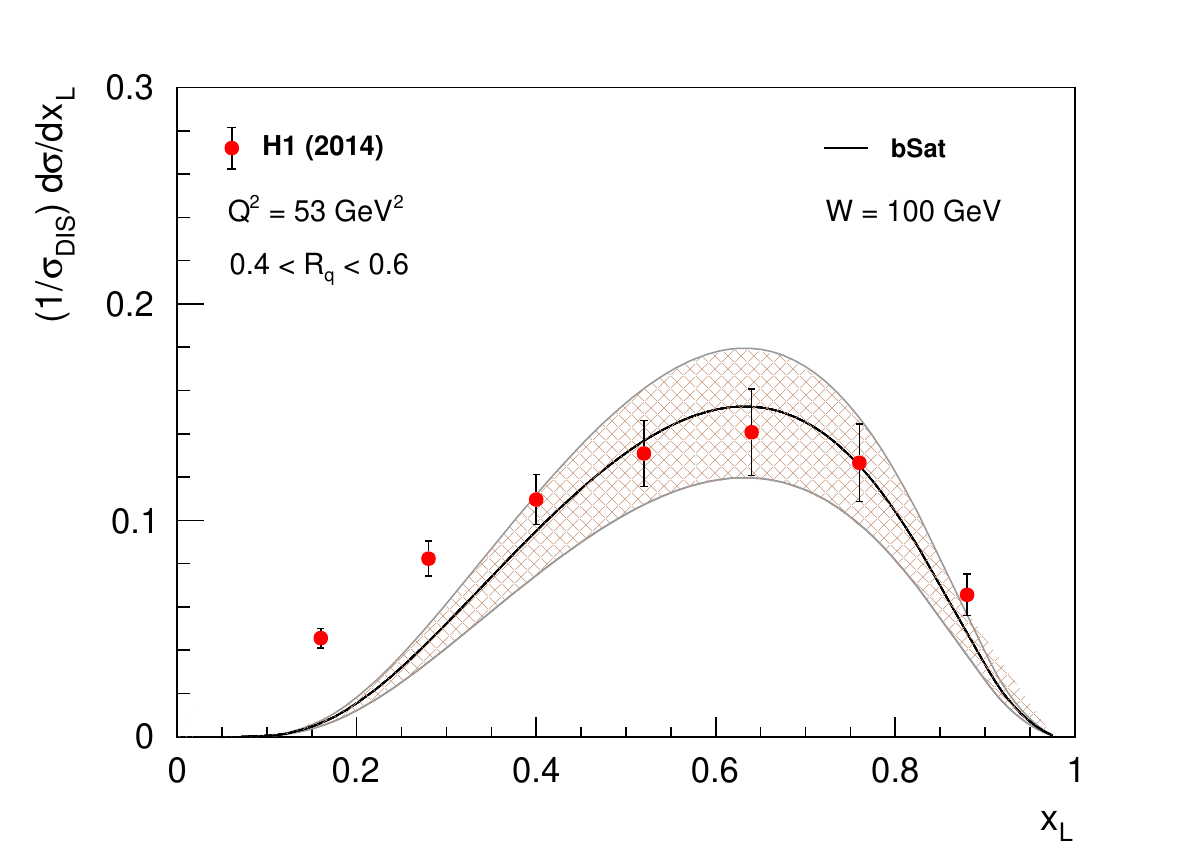}\hskip-0.1cm
	\includegraphics[width=0.5\linewidth]{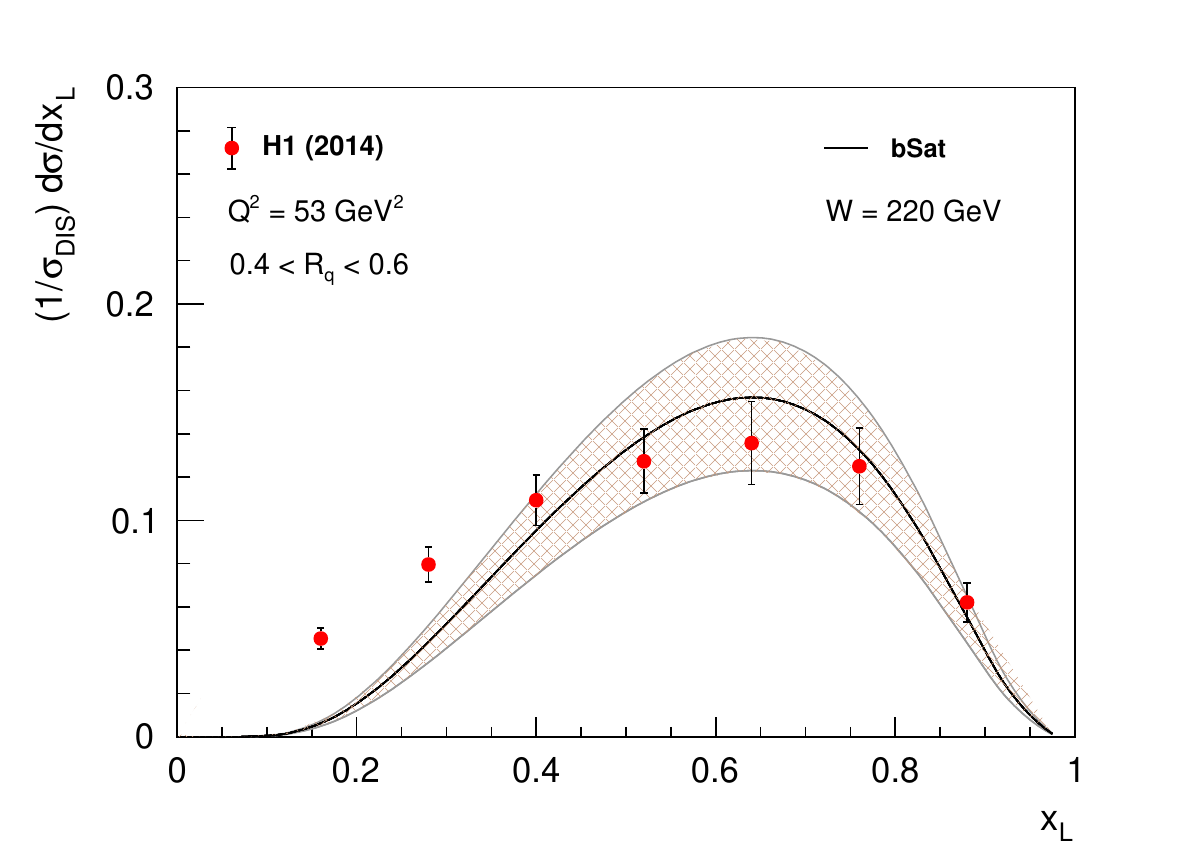}\hskip-0.1cm
	\includegraphics[width=0.5\linewidth]{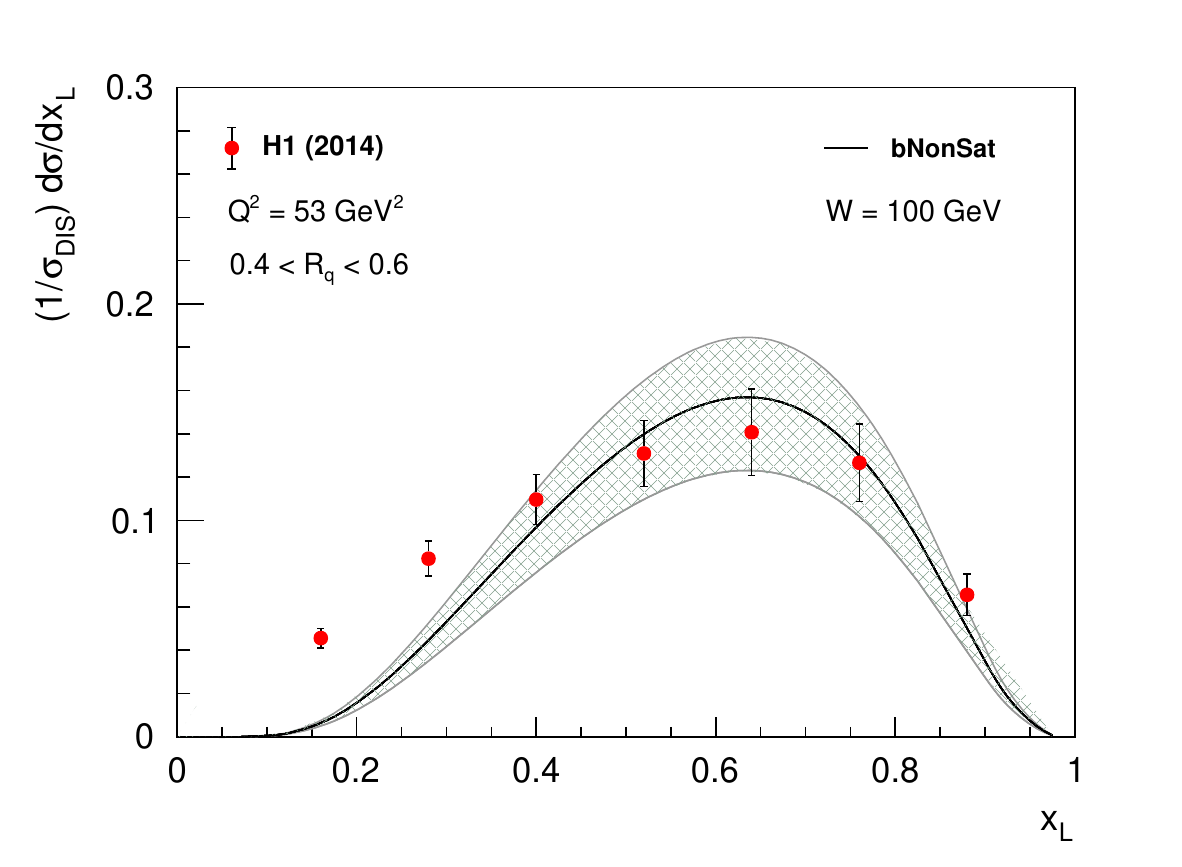}\hskip-0.1cm
	\includegraphics[width=0.5\linewidth]{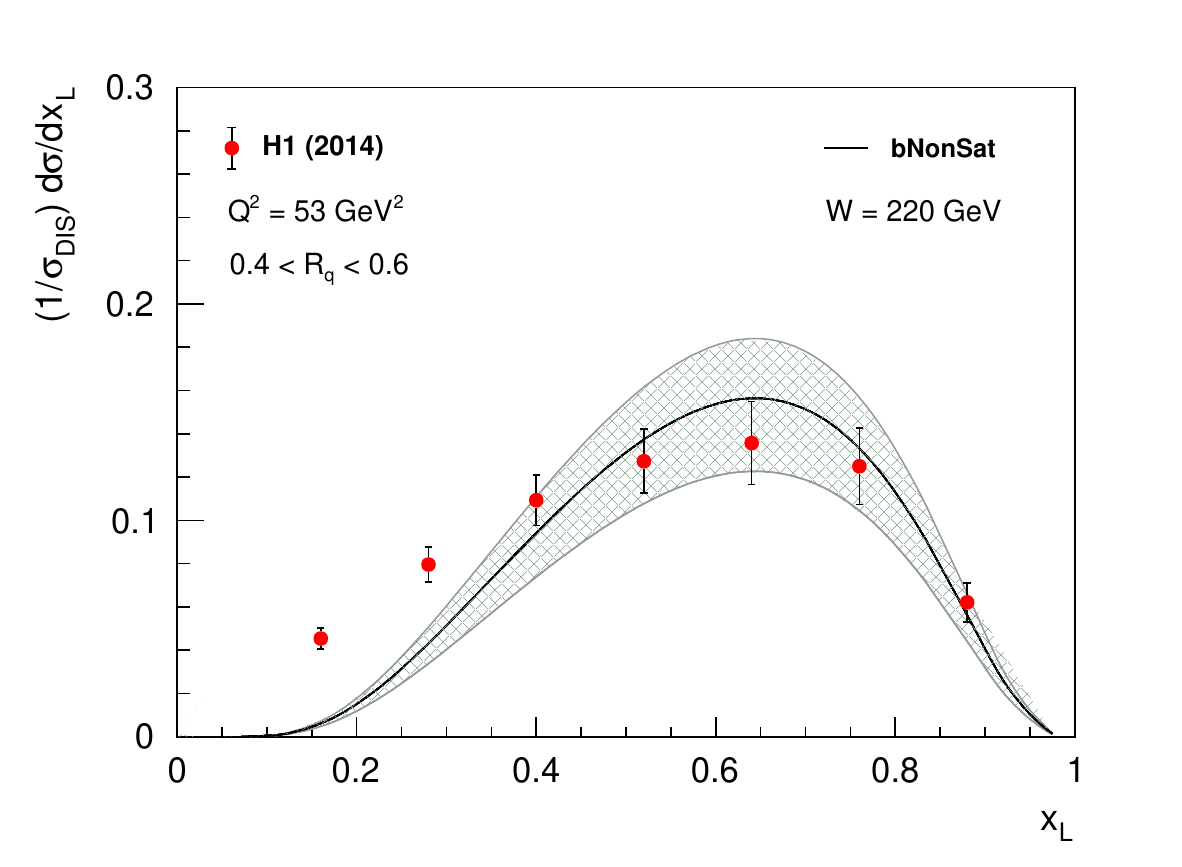}\hskip-0.1cm
	\caption{Leading neutron spectrum in bSat (first row) and bNonSat (second row) dipole models. The central line in the curves corresponds to the mean value of $R_q = 0.51$ . The model predictions are compared with the experimental data taken from H1 from Ref.\cite{H1:2014psx}.   }
	\label{Rq_spectra}
\end{figure*}
 In Fig.~\ref{Rq_spectra}, we show the leading neutron cross section as a function of $x_L$ in the dipole models with and without saturation with the uncertainty band corresponding to different values of $R_q$. We vary the parameter $0.4\leq R_q\leq 0.6$ with the central value $R_q=0.5$. We see that $R_q = 0.5$ describes the data reasonably well, which validates the assumption made in eq.~\eqref{eq:scaling}. For the rest of the paper we will use this value of $R_q$. We note that even with different values of $R_q$, the models underestimate the data at low $x_L$. In this region, the process with direct dissociation of protons into neutrons and other sub-leading processes such as $\rho$ and $a_2$ emissions contribute \cite{ZEUS:2002gig}, hence our models underestimates the cross sections for small $x_L$. The model predictions are compared with the H1 data for the two different sets where  $70<W < 100~$GeV and $190<W < 245~$GeV with $6<Q^2 < 100~$GeV$^2$ from \cite{H1:2014psx}. For the results shown here we choose the mean values of $W =100~$GeV,  $Q^2 = 53~$GeV$^2$, and  $W = 220~$GeV, $Q^2 = 53~$GeV$^2$, respectively. This choice does not have a large effect as there is a scaling with respect to $W$ and $Q^2$ in the leading neutron spectrum as discussed next.
 
\begin{figure*}[h]
	\centering
	\includegraphics[width=0.5\linewidth]{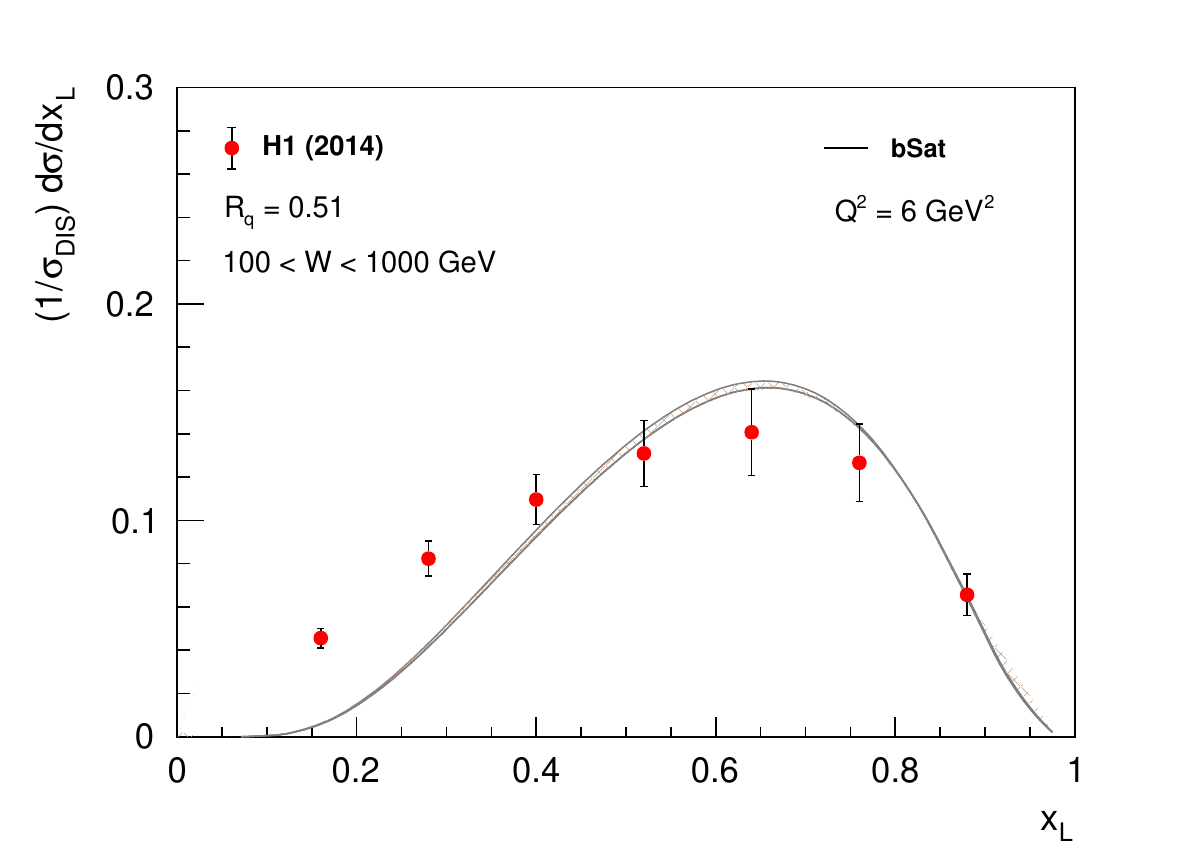}\hskip-0.1cm
	\includegraphics[width=0.5\linewidth]{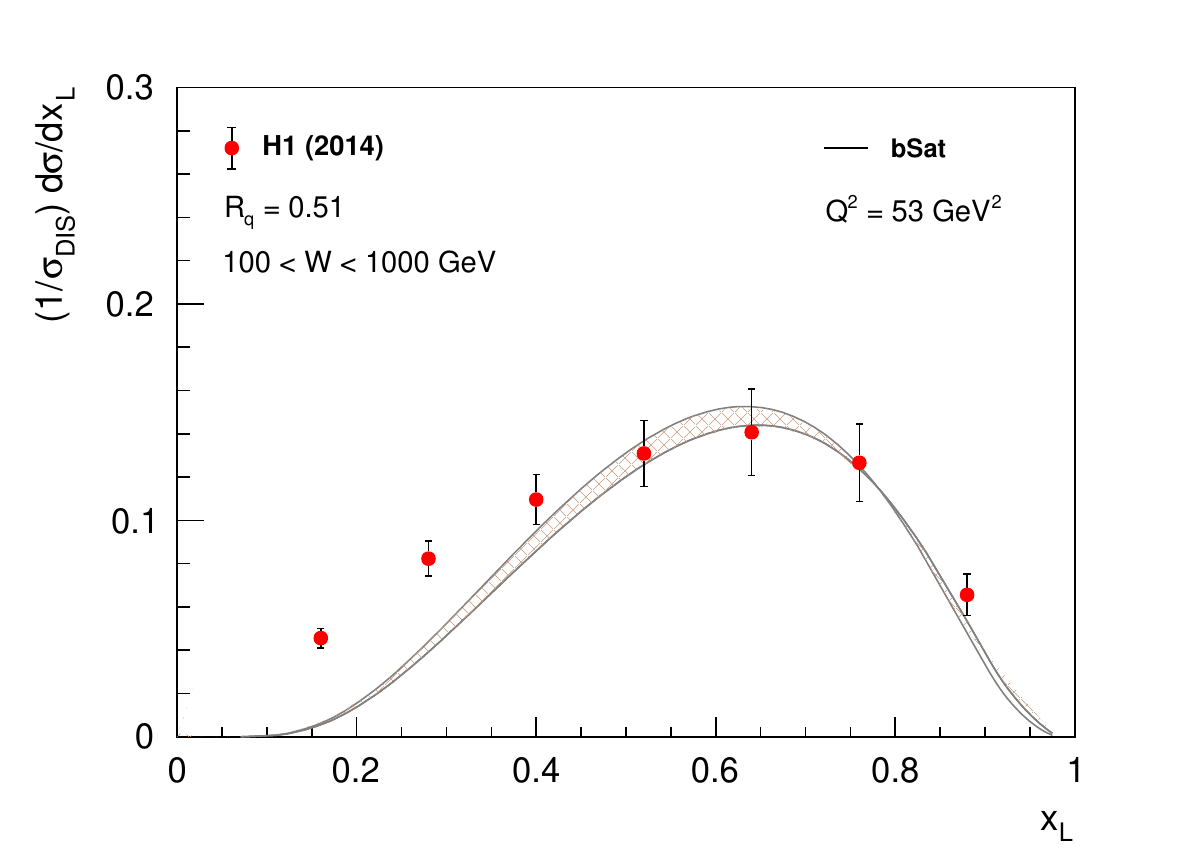}\hskip-0.1cm
	\includegraphics[width=0.5\linewidth]{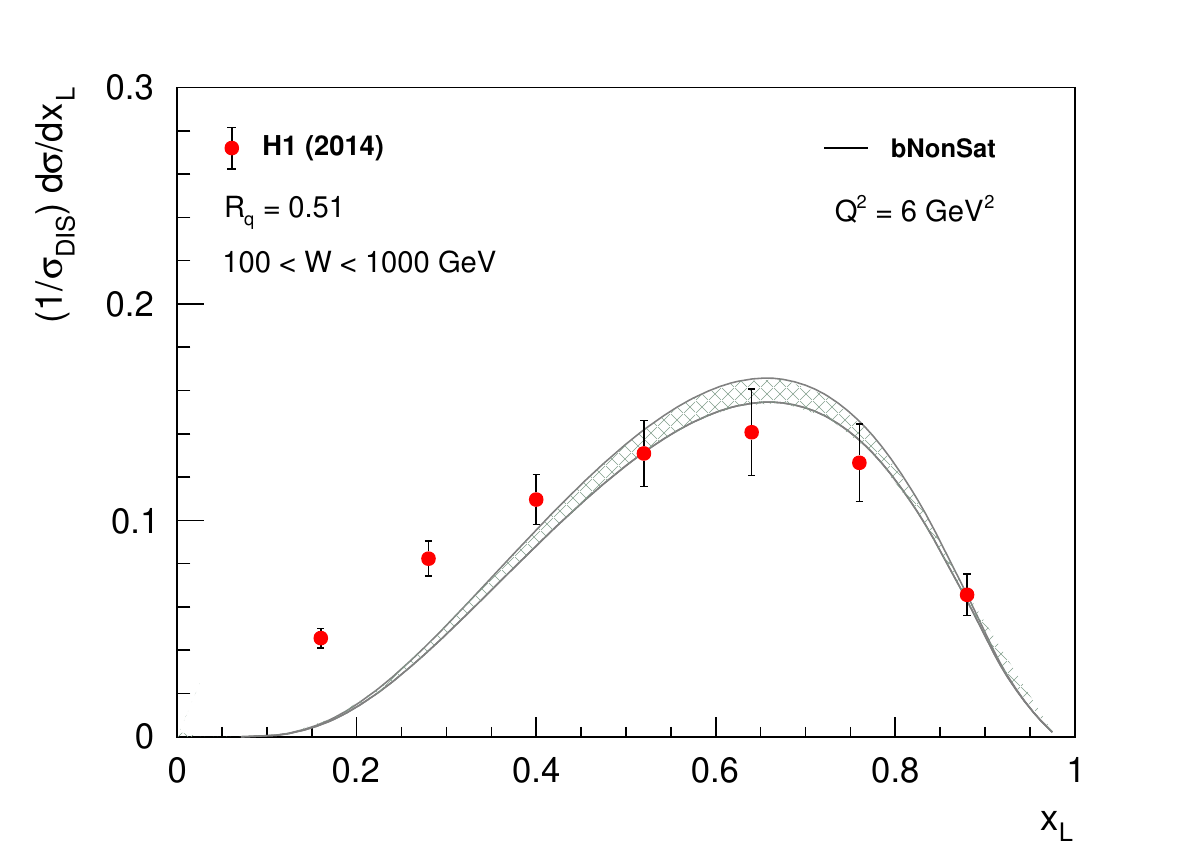}\hskip-0.1cm
	\includegraphics[width=0.5\linewidth]{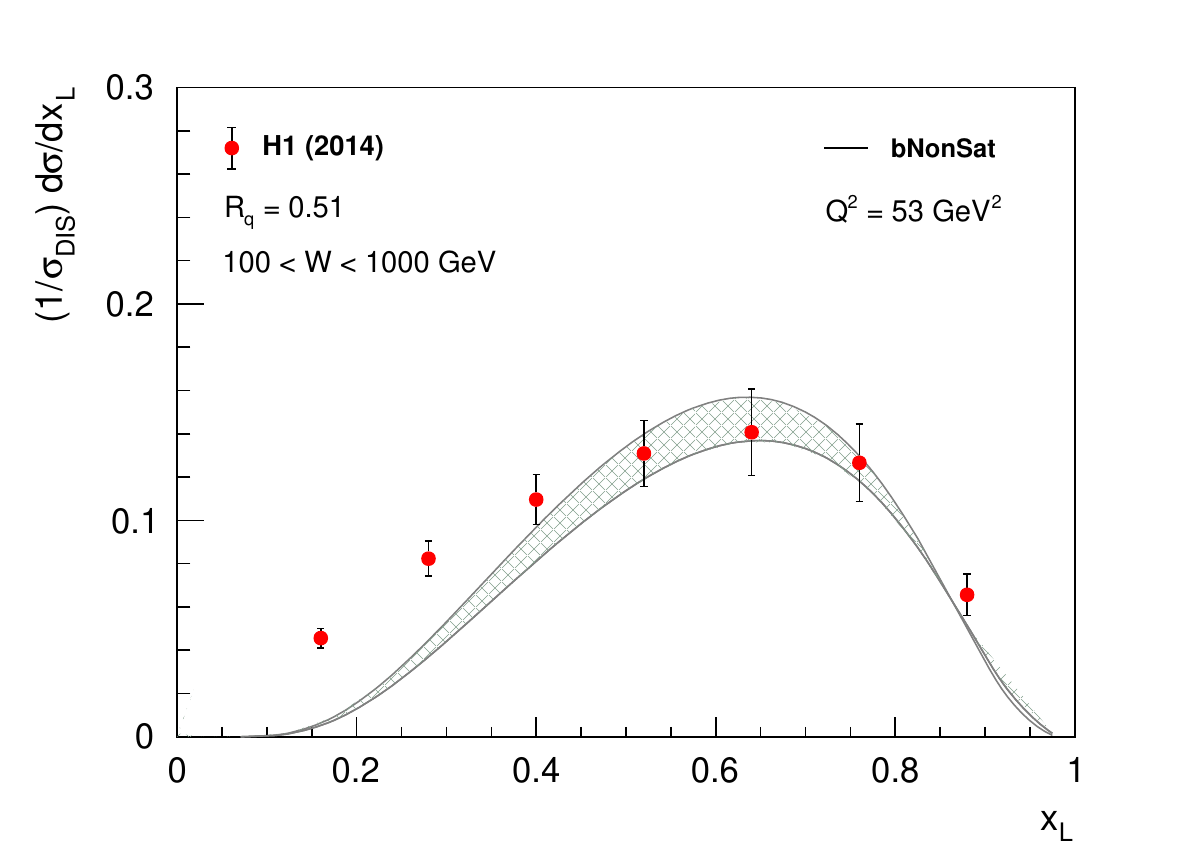}\hskip-0.1cm
	\caption{Scaling $w.r.t$ W in leading neutron spectrum in bSat (first row) and bNonSat (second row) dipole models. The lower line of band corrsponds to $W = 1000~GeV$ and the upper line of the band corrsponds to $W = 100~GeV$.   }
	\label{W_scaling}
\end{figure*}
%Fig.~\ref{W_scaling} depicts the Feynman scaling present in the semi-inclusive leading neutron spectra i.e the normalised cross-section plotted as a function of $x_L$ is independent of the values of $W$. 
Fig.~\ref{W_scaling} shows the semi-inclusive leading neutron differential cross section with respect to $x_L$ normalised to the total DIS cross section. We observe that this data exhibits Feynman scaling with respect to $W$. 
Here, the numerator ${\rm d}\sigma/{\rm d}x_L$ increases with $W$ and so does the denominator $\sigma_{DIS}$ (which is also a function of \emph{x} and $Q^2$). As a result the ratio remains fixed. For perfect scaling we expect a single curve for all the values of $W$. Instead we observe a narrow band. 
This is due to fact that the differential cross section for neutrons, $\dint \sigma/\dint x_L$, is evaluated at a scaled Bjorken variable $\hat{x}$ while the inclusive cross section $\sigma_{DIS}$ is calculated at the usual Bjorken-\emph{x}. Hence we are not comparing the two cross sections at the same \emph{x} values. This effect becomes more prominent at large $Q^2$ which is seen in Fig.~\ref{W_scaling}. The band corresponds to the W values in the range $100<W<1000~$GeV. %The lower line of band corresponds to $W = 1000~GeV$ and the upper line of the band corresponds to $W = 100~GeV$. 
We show it for two values of $Q^2 = 6, 53~$GeV$^2$ and observe that this scaling is present in both models. This is because the photon-pion cross section has the same energy dependence as the photon-proton cross section in both the models. This scaling behaviour thus justifies the main assumption we considered in Eq.\eqref{eq:scaling}, where the $\gamma^*\pi^*$ dipole cross section is equivalent to the $\gamma^*p$ dipole cross section up to normalisation. Thus, the pion and proton structure functions have identical asymptotic behaviours. This also leads us to conclude that saturation is not associated to Feynman-scaling and is present for all $Q^2$ values in both models.

\begin{figure}
	\centering
	\includegraphics[width=0.5\linewidth]{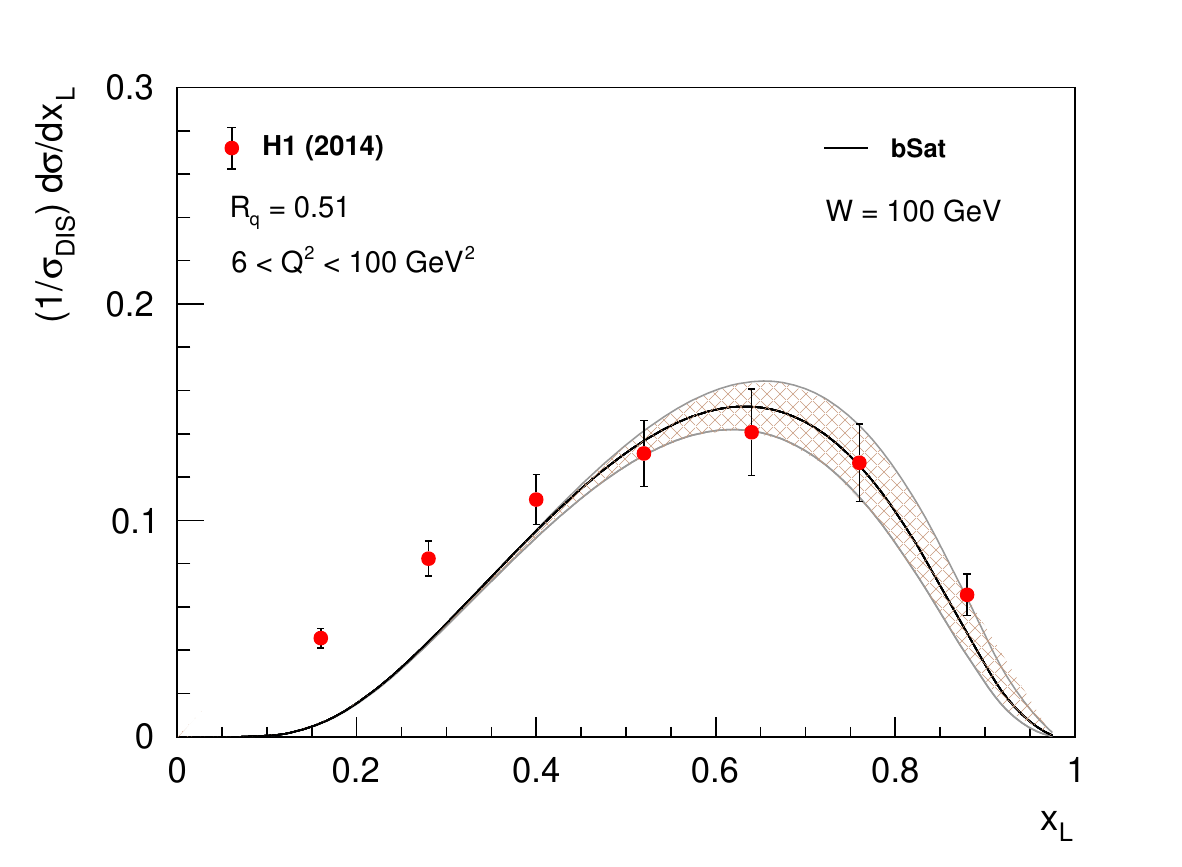}\hskip-0.1cm
	\includegraphics[width=0.5\linewidth]{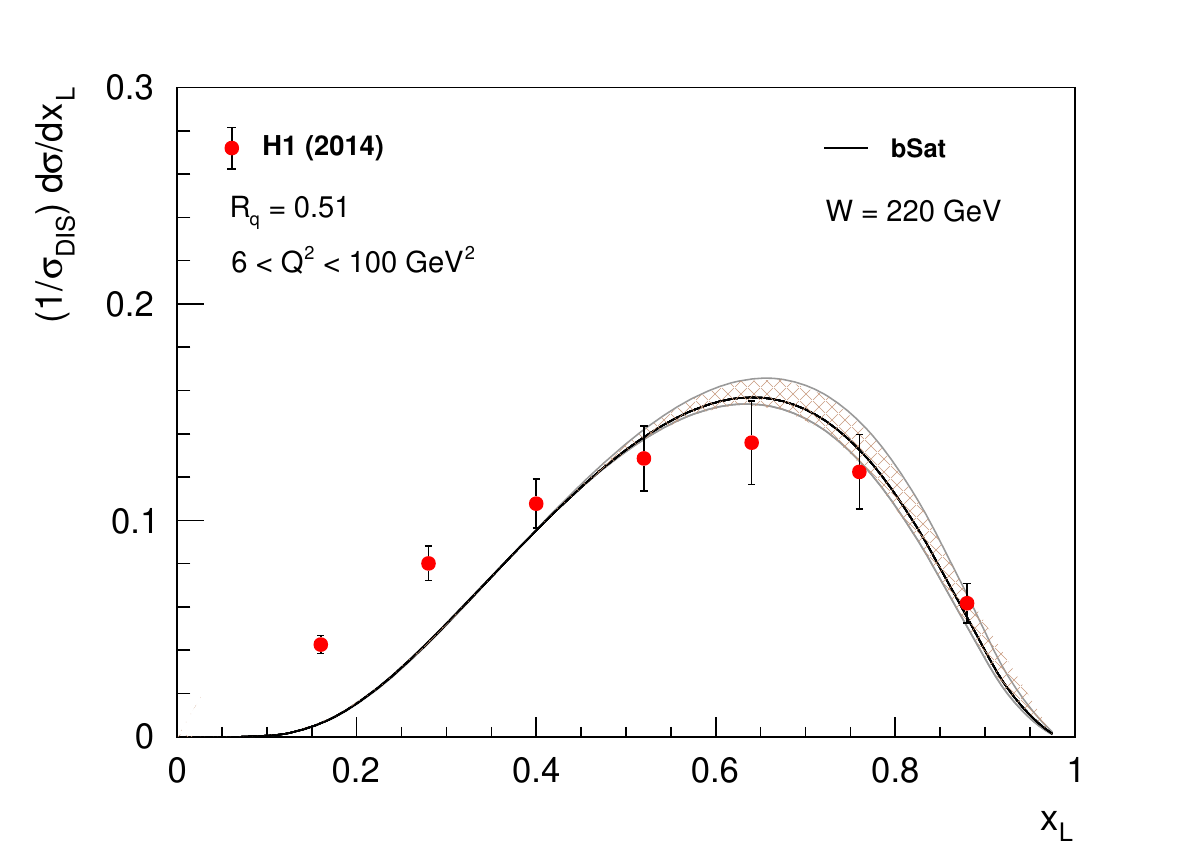}\hskip-0.1cm
	\includegraphics[width=0.5\linewidth]{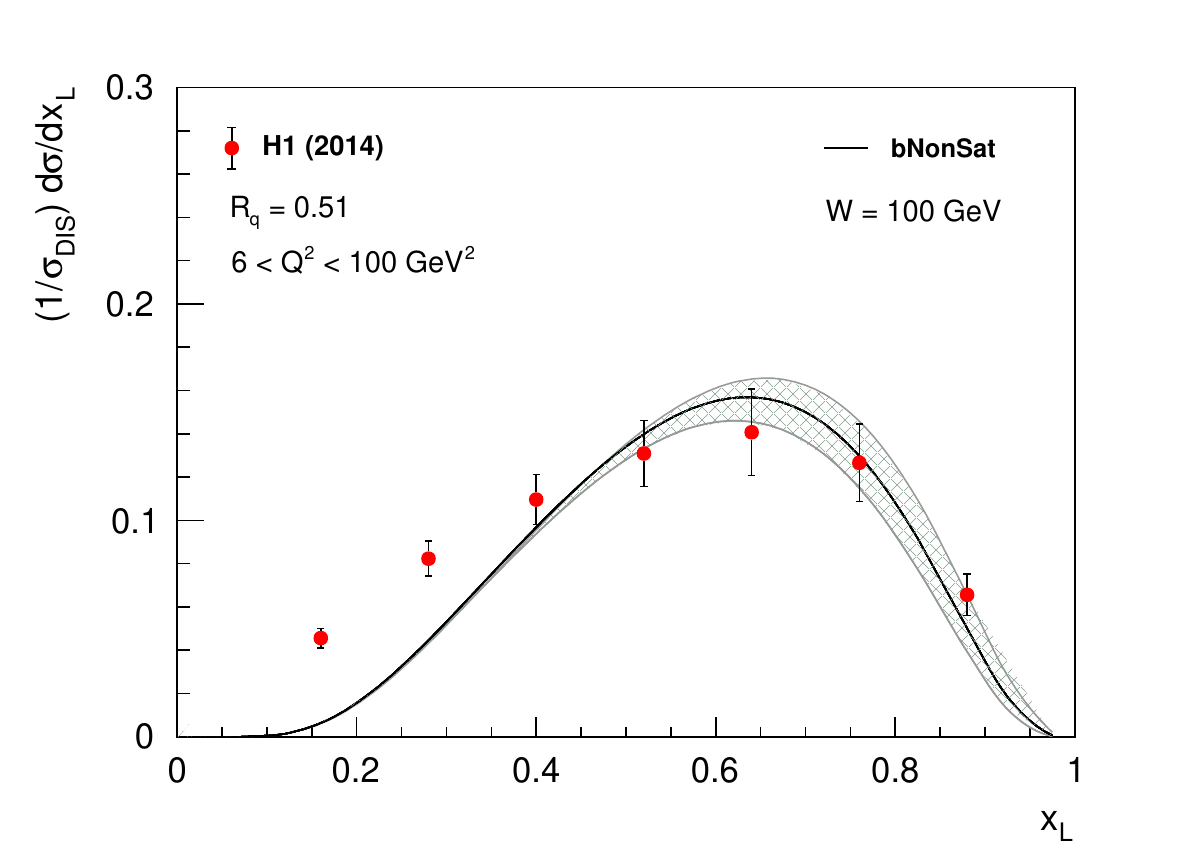}\hskip-0.1cm
	\includegraphics[width=0.5\linewidth]{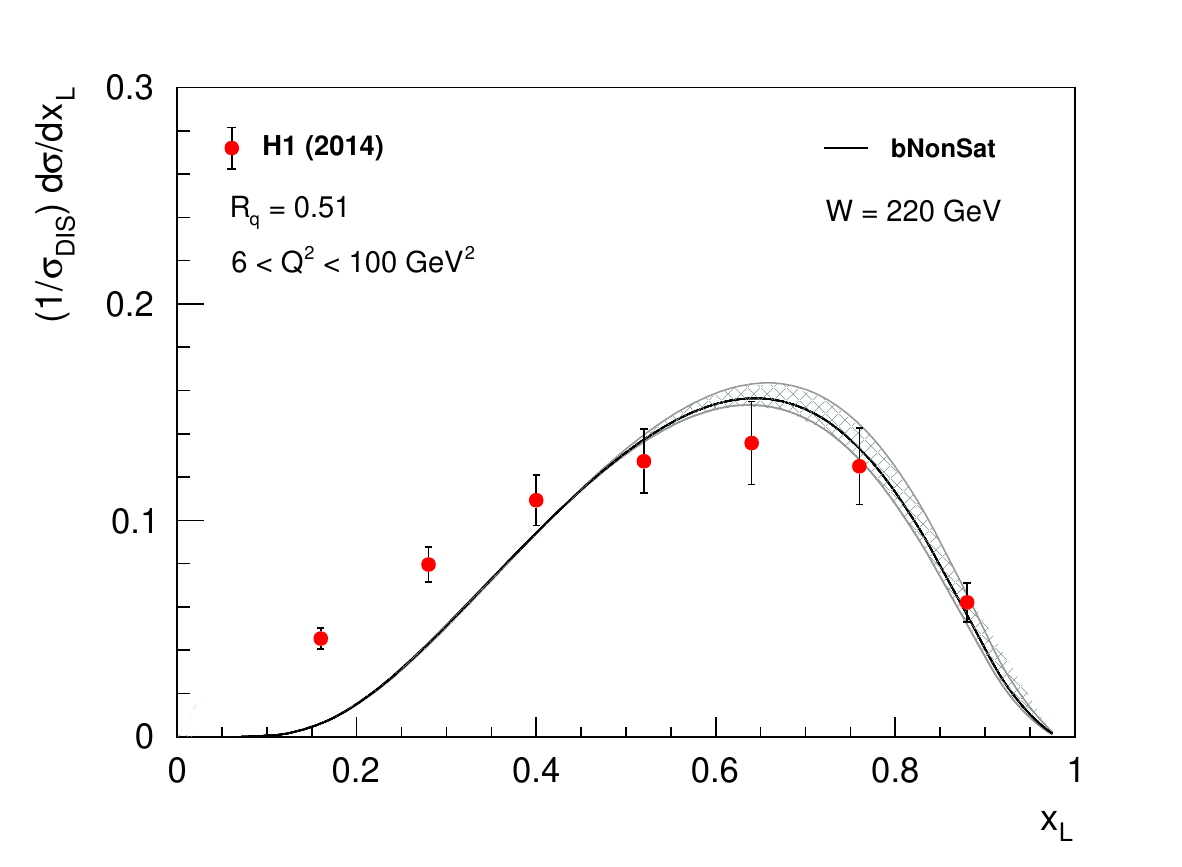}\hskip-0.1cm
	\caption{Scaling $w.r.t$ $Q^2$ in leading neutron spectrum in bSat (first row) and bNonSat (second row) dipole models. The upper line of band corrsponds to $Q^2_{min} = 6~GeV^2$ and the lower line of the band corrsponds to $Q^2_{max}=100~GeV$ and the central line corrsponds to $Q^2 = 53~GeV^2$.   }
	\label{Q2_scaling}
\end{figure}
Fig.~\ref{Q2_scaling} illustrates the scaling with respect to $Q^2$ in the leading neutron cross section. We show the normalised cross section as a function of $x_L$. The bands correspond to varying $Q^2$ in the range $6<Q^2<100~$GeV$^2$,
%The upper line of band corresponds to $Q^2 = 6~GeV^2$ and the lower line of the band corresponds to $Q^2=100~GeV^2$ 
with central line corresponding to the mean value $Q^2=53~$GeV$^2$. We show it for two values of the $W = 100, 220~$GeV. When calculating the cross section at different $Q^2$, the absorptive corrections are important, since for small $Q^2$ the dipole size is large and the dipole can re-scatter from the neutron, while for large $Q^2$, the dipole size is small and the effect of absorptive corrections dwindle. As shown in \cite{Carvalho:2020myz}, in the kinematic region considered in this study, these effects can be modelled by multiplying the flux with a factor $\mathcal{K}$ whose value vary with $Q^2$. These values are directly taken from \cite{Carvalho:2020myz} and are $\mathcal{K}= \left[0.8,0.9,1.0\right]$  for $Q^2 = \left[6,53,100\right]$ respectively. In \cite{Carvalho:2015eia}, this multiplicative factor was considered independently of $Q^2$ which resulted in violation of $Q^2$-scaling. We observe a narrow band here as well. Again, this is because the leading neutron cross section is calculated at $\hat{x}$ while the proton cross section is evaluated at $x$. This effect become more prominent at small $W$ where we cannot neglect the $Q^2$ in eq.\eqref{eq:x}. The theoretical predictions for both models are well within the experimental uncertainties.

\begin{figure}
	\centering
	\includegraphics[width=\linewidth]{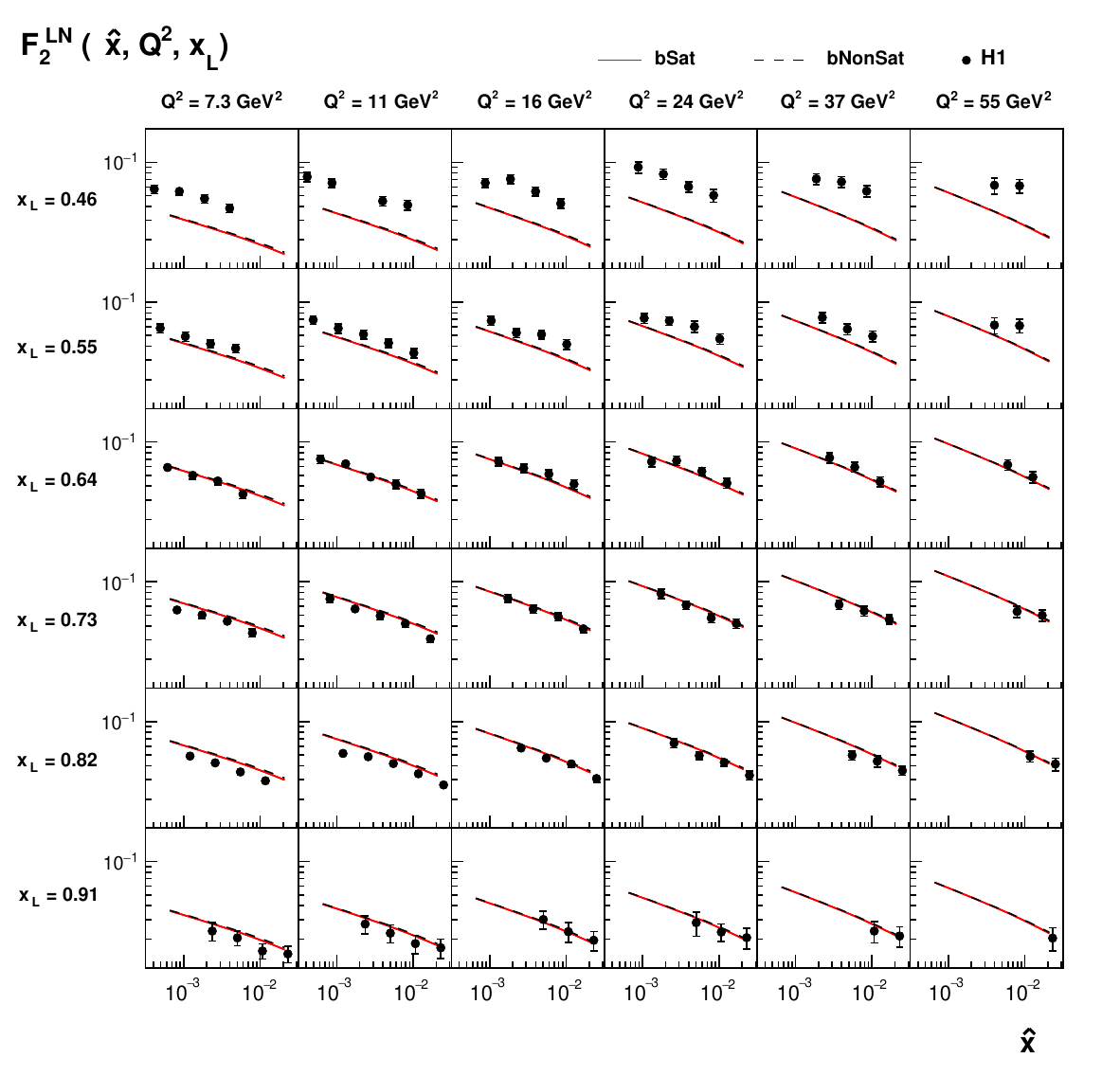}
	\caption{Leading neutron structure function $F_2^{LN}(\hat{x},Q^2,x_L)$ as function of $\hat{x}$, for different values of $Q^2$ and $x_L$, in the bSat (solid line) and the bNonSat (dotted line) dipole models. 
	}
	\label{F2_pions}
\end{figure}
In Fig.~\ref{F2_pions}, we present the predictions for the leading neutron structure function $F_2^{LN}$ as a function of Bjorken-$\hat{x}$ for different values of $x_L$ with varying $Q^2$ in the bSat and bNonSat dipole models and confront them with the HERA measurement from \cite{H1:2010hym}. Both the models provide a good description of the energy dependence of the data for $x_L\geq 0.55$, while for lower $x_L$ values the models underestimates the data. This is because the one pion exchange approximation holds good for $0.5 \lesssim x_L\lesssim 0.9$ as discussed earlier and both models give an excellent description for  $x_L\geq 0.64$. The curves for both the models with and without saturation are indistinguishable in the whole kinematic region. 

\begin{figure*}
	\centering
	\includegraphics[width=0.5\linewidth]{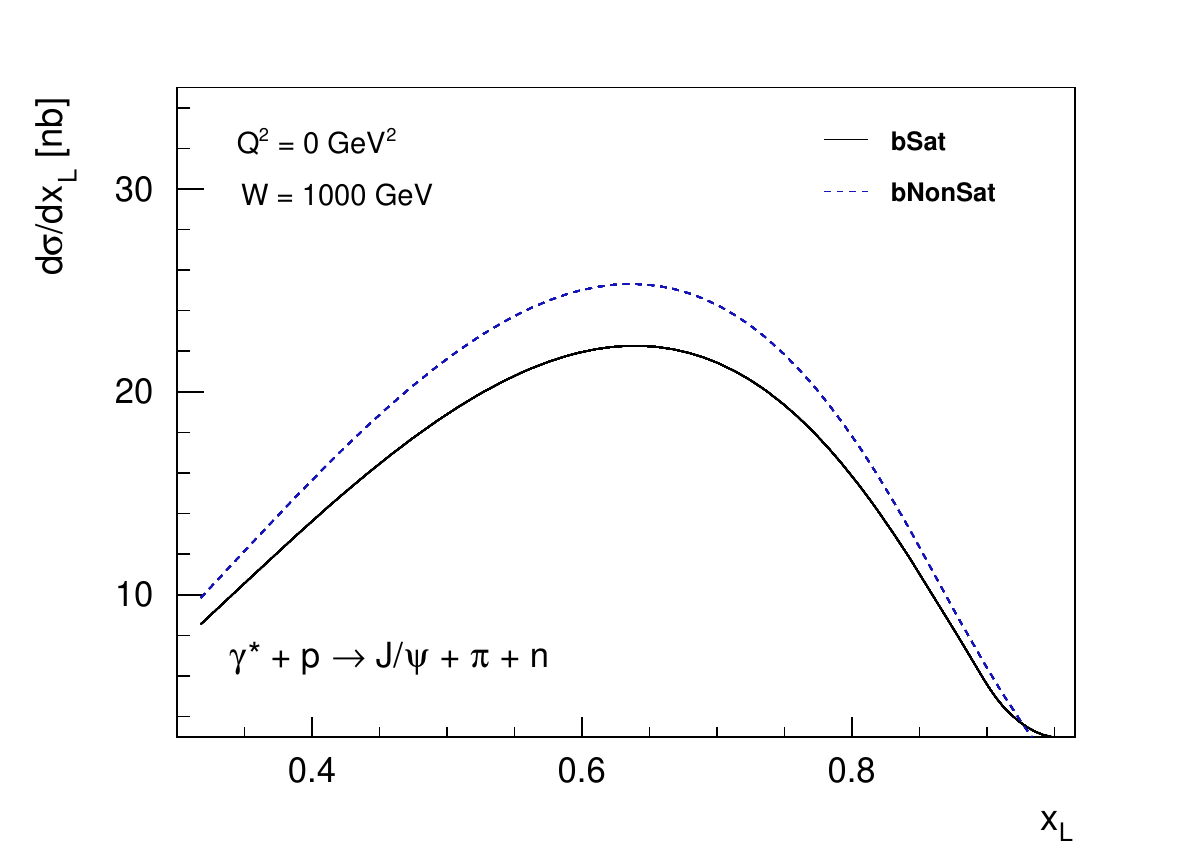}\hskip-0.1cm
	\includegraphics[width=0.5\linewidth]{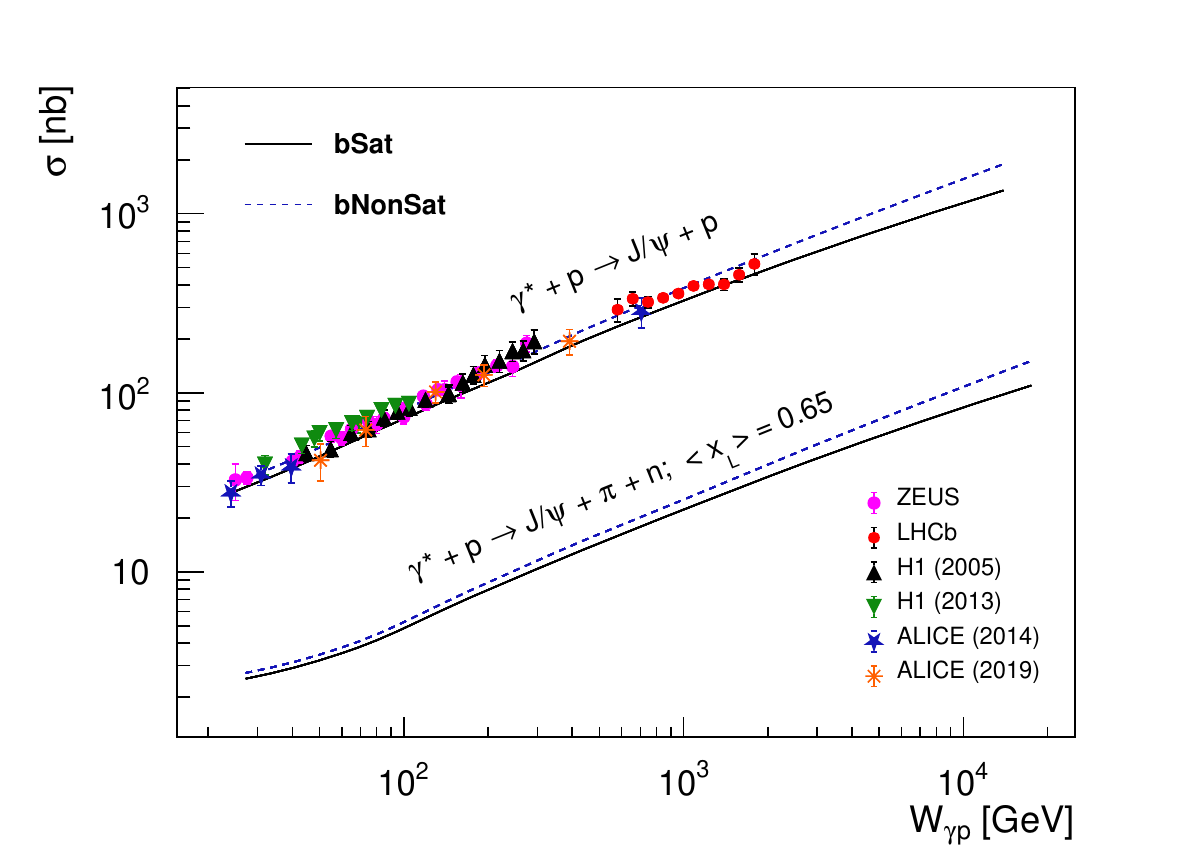}\hskip-0.1cm
	\includegraphics[width=0.5\linewidth]{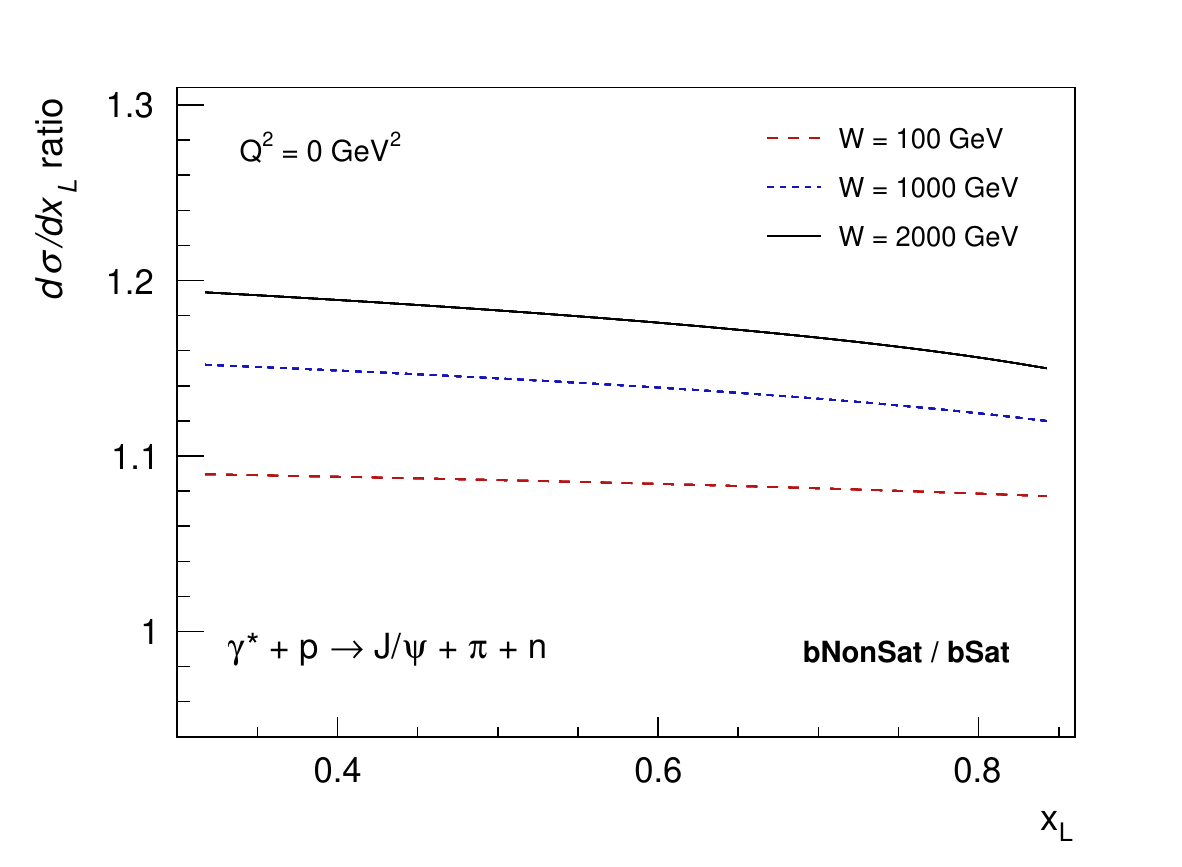}\hskip-0.1cm
	\includegraphics[width=0.5\linewidth]{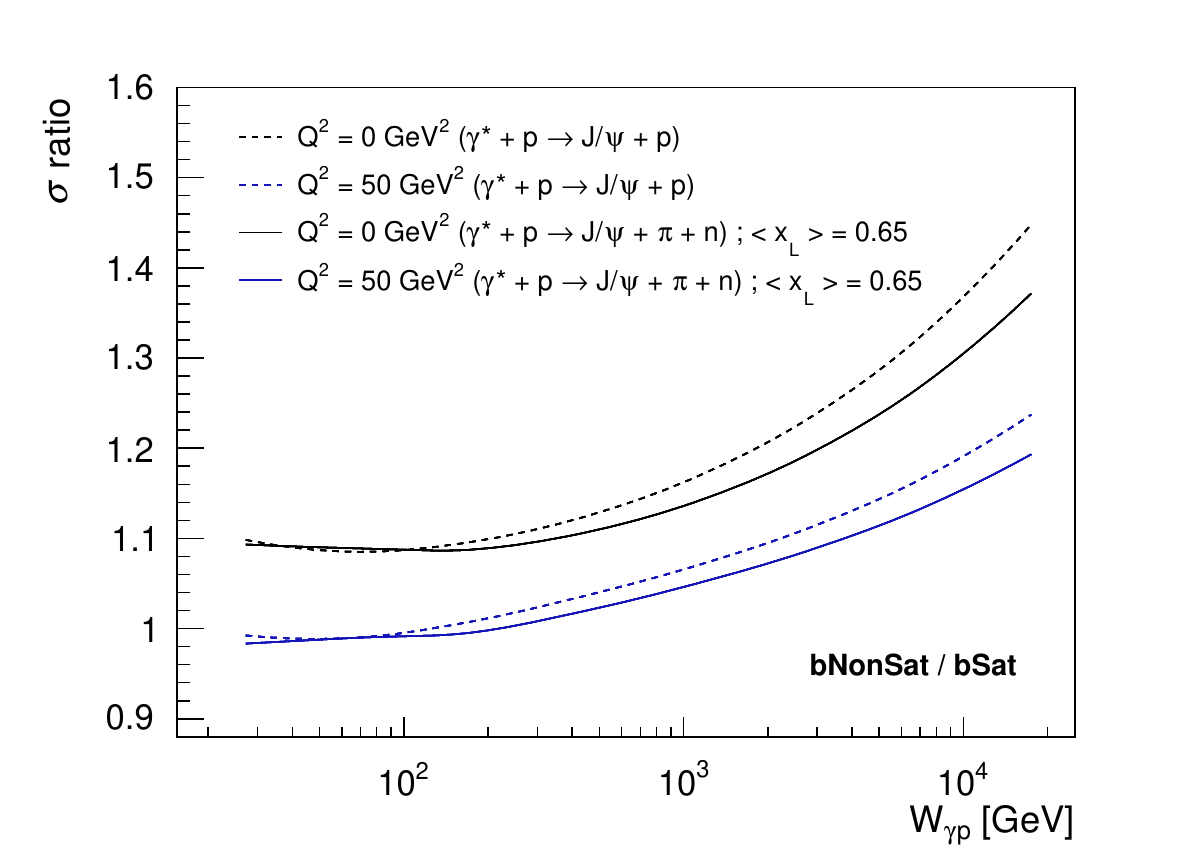}\hskip-0.1cm
	\caption{Exclusive $J/\psi$ production with leading neutrons in $ep$ collisions at future colliders in bSat model (solid, black) and bNonSat model (dashed, blue).        }
	\label{Exclusive_LN}
\end{figure*}
In Fig.~\ref{Exclusive_LN}, we plot the Feynman-\emph{x} distribution, the energy dependence, and the corresponding bNonSat to bSat cross section ratios of the differential and total cross section for the leading neutrons in exclusive $J/\psi$ production.  The absorptive corrections are not known for exclusive diffraction but due to the large mass of $J/\psi$, the dipole size is small and the absorptive effects will be suppressed. We include the absorptive effects in calculating the spectrum by multiplying the flux with $\mathcal{K} = 0.8$. Moreover, while calculating the ratios of bNonSat and bSat cross sections this effect is nullified. In the first row, in Fig.~\ref{Exclusive_LN}, we present the predictions for the differential cross section with respect to $x_L$ and the energy dependence of the total cross section for bSat and bNonSat model and we observe that the saturation effects suppress the cross section. The difference between the models increase with $W$, and is larger for small $Q^2$. This is seen clearly in the ratio plots, where for very large $W$ and small $Q^2$ the non-saturated model is 50-60\% larger than the saturated model. The ratio plot also shows that the saturation effect is nearly independent of $x_L$. We also show the energy dependence and the ratio of bNonSat to bSat cross sections for the proton case and we see that though the exclusive leading neutron spectrum is sensitive to non-linear effects they are less so than that of the proton. 
%{\comment Conclusion} Hence we conclude that the leading neutron cross section is not a smoking gun signature for the saturation. 

\begin{figure}
	\centering

		\includegraphics[width=0.448\linewidth]{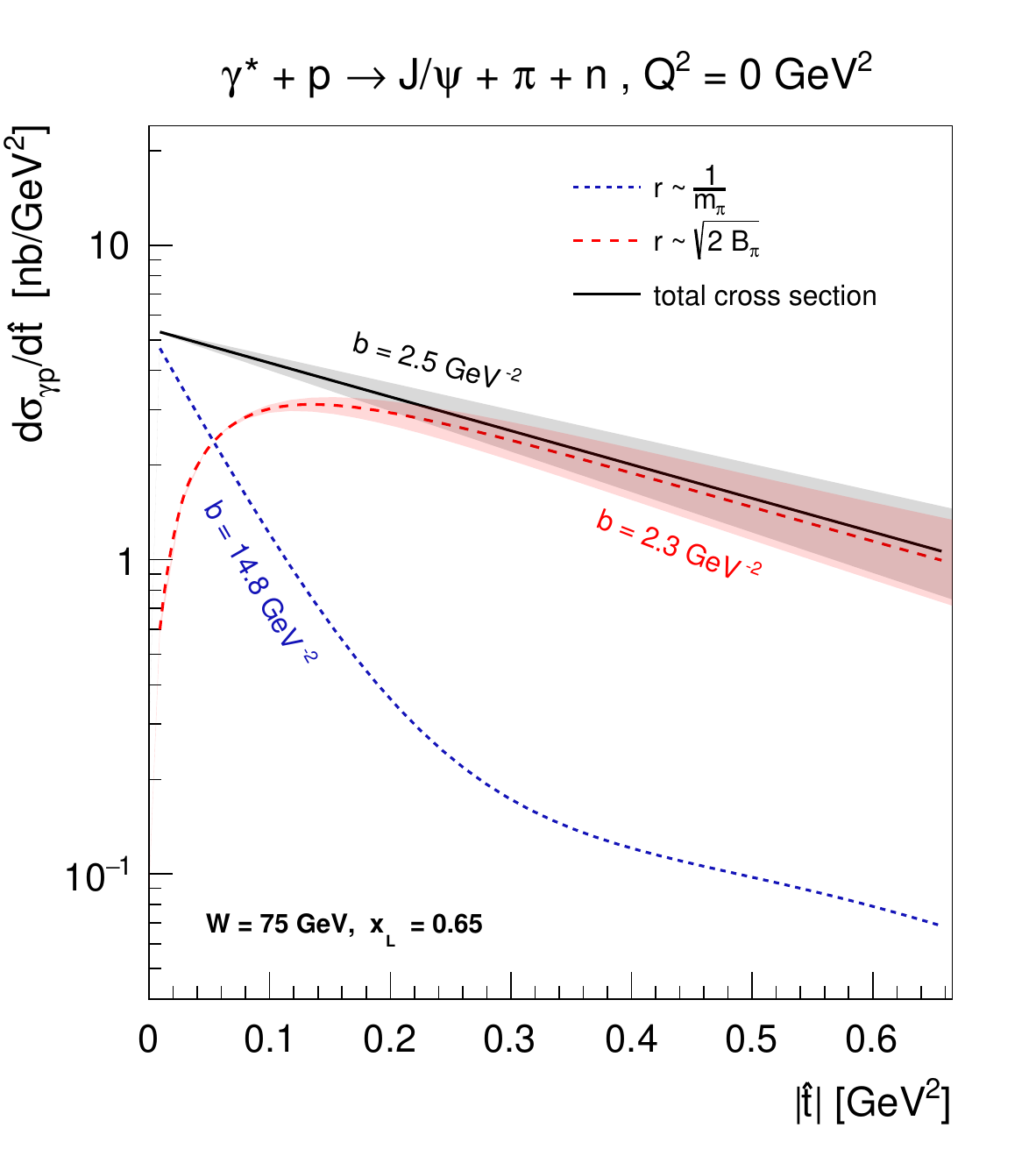}\hskip1cm
		\includegraphics[width=0.447\linewidth]{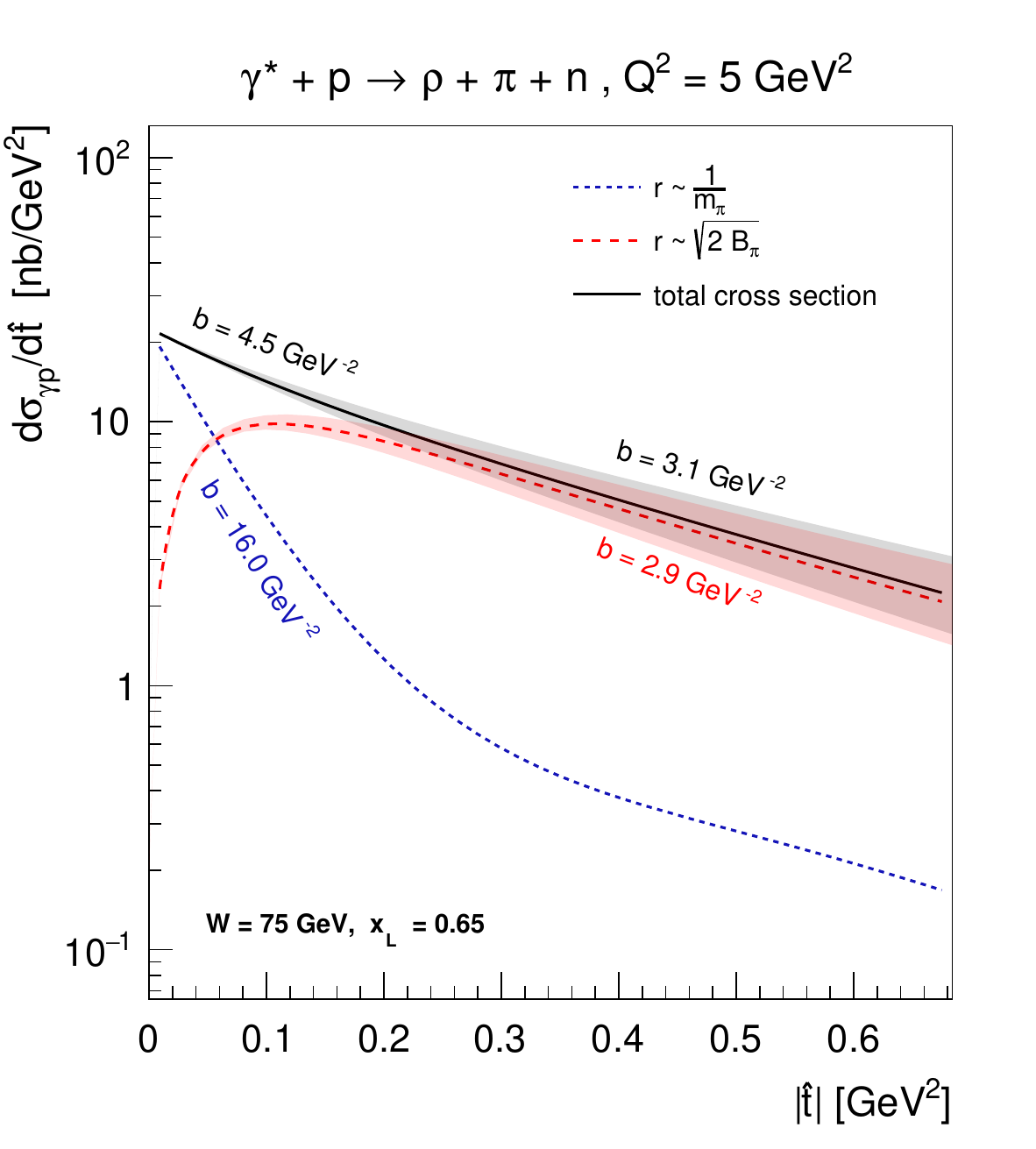}
	\caption{${\hat t}$-dependence of the exclusive J/$\psi$ (left) and $\rho$ (right) production with leading neutron in $\gamma^*p$ scattering in bNonSat model. The band corresponds to $B_{\pi} = 2.0 \pm 0.5$~GeV$^{-2}$. The $b$ values are extracted by fitting $d\sigma/dt \propto \text{exp}(-b |t|)$ for the different curves. }
	\label{tDep}
\end{figure}
In Fig.~\ref{tDep} we show our prediction for the $\hat t$ spectrum for an exclusive $J/\psi$ (left) and $\rho$ (right) meson produced through the interaction with a virtual pion cloud. Since we calculate the virtual pion distribution with Yukawa theory, the only parameter in this model is $B_\pi$, the width of the gluon distribution in the real pion. There is no available data which directly restricts this parameter. However, there are a few studies that give us some clues. Firstly, we note that the charge radius of the pion is measured to be $r_\pi=0.657\pm 0.003$ fm \cite{Ananthanarayan:2017efc}. If we assume that the gluon to charge radius ratio is the same in pions and protons, we get $B_\pi=r_\pi^2/r_p^2 B_p= (0.657/0.840)^2\cdot 4~$GeV$^{-2}\approx 2.44$~GeV$^{-2}$ \cite{Gao:2021sml}. The pion gluon radius is also extracted from the Belle measurements \cite{Belle:2012wwz,Belle:2015oin} at KEKB in \cite{Kumano:2017lhr} using the Generalized Distribution Amplitudes in hadron-pair production in a two-photon process $\gamma^* \gamma \rightarrow \pi^0 \pi^0$  and was found to be $B_\pi=1.33-1.96~$GeV$^{-2}$. Also, H1 measured the $\hat t$ spectrum for exclusive $\rho$ photo-production with leading neutrons in $ep$ scattering \cite{H1:2015bxa}, as this process lacks a hard scale we are not able to make a direct comparison, but this spectrum suggests $B_\pi\approx 2.3$~GeV$^{-2}$. We therefore present our results with bands for $B_\pi=2\pm0.5~$GeV$^{-2}$. The experimentally observed cross section for such events is the total cross section plotted in fig.~\ref{tDep} and in our framework this cross section consists of two different slopes; the first corresponds to the virtual pion interaction at small $|{\hat t}|$ and second due to interaction with the real pion at moderate $|{\hat t}|$. We see that in $J/\psi$ production the resulting curve has a similar slope over the whole spectrum, while in $\rho$ production one can distinguish the two slopes a bit clearer. 
%The ${\hat t}$ spectrum in Fig.~\ref{tDep} demonstrates the presence of these two slopes and this feature will be further pronounced for $\rho$ production at low $Q^2$ values where the large dipole size can probe the full radial extent of the pion cloud. We have strong indications of the validation of our model from 
The $\hat t$ spectrum measured by H1 \cite{H1:2015bxa} clearly exhibits these two distinct slopes.

%%%%%%%%%%%%%%%%%%%%%%%%%%%%%%%%%%%%%%%%%%%%%%%%%%%%%%%%
\section{Conclusions and Discussion}
We have investigated virtual photon scattering with the pion cloud of protons in $ep$ scattering using two versions of the impact parameter dependent dipole model, with and without saturation effects. We have assumed that up to normalisation, the pion's structure is equivalent to the proton's at small $x$, which we have demonstrated holds good within the precision of HERA measurements. We also show that both models describe the measured $F_2^{LN}$ well (for $x_L>0.5$). 
More precise measurement of leading neutron processes, for example from the EIC, would be able to further substantiate (or reject) the assumption of small-$x$ hadron universality.

%Moreover, the Mandelstam $t$-spectra measurements of exclusive $J/\psi$ vector meson production with leading neutrons would be able to tell us the difference in the geometrical distribution of gluons in pions compared to protons. 
We have also investigated claims that so called Feynman scaling is a consequence of saturation. Feynman scaling appears in the ratio of the differential leading neutron cross section with respect to $x_L$ to the total DIS cross section. We found that Feynman scaling holds as a function of both $Q^2$ and $W$ in both the saturated and unsaturated models, and thus is independent of non-linear effects. The measured $F_2^{LN}$ data also do not exhibit any saturation effect. Exclusive $J/\psi$ production is more sensitive to non-linear effects as its cross section depends on the square of the gluon density. Here, there is a clear difference between the model predictions for large $W$. However, these non-linear effects are smaller in leading neutron $\gamma^*\pi^*$ processes than in $\gamma^*p$ processes. This is to be expected, as the pion is probed at larger momentum fractions than the proton. %We showed that the ratio of the two models as a function of $x_L$ is constant, which further indicates that Feynman scaling is not a saturation effect. 
The universality of this latter process between pions and protons is yet experimentally untested. 

%We show that the dipole model can make predictions for $F_2^\pi$. In this paper we use a single parameter $R_q$ which is the difference in normalisation between the dipole cross sections for pions and protons. In principle we could instead perform an independent fit of the parameters of the dipole model to $F_2^\pi$ measurements, which would be preferable in the future. 
We expect the universality between pions and protons to break down when measuring the $\hat t$ spectrum, as this is sensitive to the spatial distribution of gluons in the struck hadron. We have shown how to calculate the spatial gluon structure in virtual pions with the dipole model using a model where the virtual pion wave-function is given by Yukawa theory, and at larger $|\hat{t}|$ where we resolve the pion in the pion cloud, we need to consider event-by-event fluctuations in order to correctly predict the $|\hat{t}|$ spectrum. We presented our resulting predictions. We see that the cross section is large enough to show up as part of the incoherent $t$-spectrum in $\gamma^*p$ measurements. However, these measurements at HERA have hitherto excluded events with a forward neutron and pion and we can therefore not see the pion cloud contribution in the $ep$ incoherent cross section measurements. In principle this would be included in the incoherent AA cross section in ultra-peripheral collisions (UPC) at RHIC and LHC, where the final state contains a nucleus with the same A but with $Z\pm1$, which would subsequently break up, and the breakup remnants can be measured by the forward detectors such as a zero degree calorimeter. However, the cross section presented here is too small to be visible in these events. It should be possible to extract this cross section in $J/\psi$ production from existing HERA data, as well as from UPC events at RHIC and LHC with at least one proton in the initial state.

We plan to extend this study to $e$A collisions at the EIC. It is an open question how different the pion clouds in heavy nuclei are from those of the constituent protons and neutrons. There are no measurements of these effects at small $x$. Depending on how well the EIC will be able to tag and id the final state pions and/or the transformed nucleus, it could be able to measure the pion clouds of protons and neutrons separately.  

\section*{Acknowledgements}
 The work of A. Kumar is supported by the Department of Science \& Technology, India under Grant No. DST/INSPIRES/03/2018/000344. We thank all the members of the HEP-PH group and the Physics Department of IIT Delhi.

%\section*{References}
\bibliographystyle{elsarticle-num}
\bibliography{mybibfile}
\biboptions{sort&compress}
\end{document}